\documentclass[%
reprint,
superscriptaddress,
w%amsmath,
amssymb,amsmath,
prl,
]
{revtex4-2}
\usepackage[english]{babel} %francais, polish, spanish, ...
\usepackage [table]{xcolor}
\usepackage{graphicx, upgreek, subfigure} %%For loading graphic files
\usepackage{float}
\usepackage{textcomp}
\usepackage{graphicx}% Include figure files
\usepackage{dcolumn}% Align table columns on decimal point
\usepackage{bm}% bold math
\usepackage[mathlines]{lineno}% Enable numbering of text and display math
\usepackage{scalerel,amssymb}
\usepackage{physics}
\usepackage{array}

\usepackage{lineno}
\newcommand{\SI}{SI}
\def\msquare{\mathord{\scalerel*{\Box}{gX}}}
\begin{document}

%\begin{abstract}
%Superconducting qubits Quantum memories are a critical part of many 
%\end{abstract}

\title{Random-access quantum memory using chirped pulse phase encoding}

\author{James~O'Sullivan}
\altaffiliation{These authors have contributed equally to this work} 
\affiliation{London Centre for Nanotechnology, UCL, 17-19 Gordon Street, London, WC1H 0AH, UK}
\author{Oscar~W.~Kennedy}
\altaffiliation{These authors have contributed equally to this work} 
\affiliation{London Centre for Nanotechnology, UCL, 17-19 Gordon Street, London, WC1H 0AH, UK}
\author{Kamanasish Debnath}
\affiliation{Department of Physics and Astronomy, Aarhus University, DK-8000 Aarhus C, Denmark}
\author{Joseph Alexander}
\affiliation{London Centre for Nanotechnology, UCL, 17-19 Gordon Street, London, WC1H 0AH, UK}
\author{Christoph~W.~Zollitsch}
\affiliation{London Centre for Nanotechnology, UCL, 17-19 Gordon Street, London, WC1H 0AH, UK}
\author{Mantas~\v{S}im\.{e}nas}
\affiliation{London Centre for Nanotechnology, UCL, 17-19 Gordon Street, London, WC1H 0AH, UK}
\author{Akel Hashim}
\affiliation{Quantum Nanoelectronics Laboratory, Department of Physics, UC Berkeley, California 94720, USA}
\affiliation{Lawrence Berkeley National Laboratory, Berkeley, CA 94720, USA}
\author{Christopher~N.~Thomas}
\affiliation{Cavendish Laboratory, University of Cambridge, JJ Thomson Ave,  Cambridge CB3 0HE, UK}
\author{Stafford Withington}
\affiliation{Cavendish Laboratory, University of Cambridge, JJ Thomson Ave,  Cambridge CB3 0HE, UK}
\author{Irfan Siddiqi}
\affiliation{Quantum Nanoelectronics Laboratory, Department of Physics, UC Berkeley, California 94720, USA}
\affiliation{Lawrence Berkeley National Laboratory, Berkeley, CA 94720, USA}
\author{Klaus M\o{}lmer}
\affiliation{Department of Physics and Astronomy, Aarhus University, DK-8000 Aarhus C, Denmark}
\author{John~J.~L.~Morton}
\email{jjl.morton@ucl.ac.uk}
\affiliation{London Centre for Nanotechnology, UCL, 17-19 Gordon Street, London, WC1H 0AH, UK}
\affiliation{Department of Electrical and Electronic Engineering, UCL, Malet Place, London, WC1E 7JE, UK}

% \maketitle

\begin{abstract} 
    As in conventional computing, memories for quantum information benefit from high storage density and, crucially, random access, or the ability to read from or write to an arbitrarily chosen register.
    However, achieving such random access with quantum memories in a dense, hardware-efficient manner remains a challenge. %, for example requiring dedicated cavities per qubit or pulsed field gradients. 
    Here we introduce a protocol using chirped pulses to encode qubits within an ensemble of quantum two-level systems, offering both random access and naturally supporting dynamical decoupling to enhance the memory lifetime. 
    We demonstrate the protocol in the microwave regime using donor spins in silicon coupled to a superconducting cavity, storing up to four multi-photon microwave pulses in distinct memory modes and retrieving them on-demand up to 2~ms later. 
    %A further advantage is the natural suppression of superradiant echo emission, which we show is critical when approaching unit cooperativity. 
    This approach offers the potential for microwave random access quantum memories with lifetimes exceeding seconds, while the chirped pulse phase encoding could also be applied in the optical regime to enhance quantum repeaters and networks.
\end{abstract}

\maketitle

Quantum memories (QMs) capable of faithfully storing and recalling quantum states on-demand are powerful ingredients in building quantum networks~\cite{kimble2008quantum} and quantum processors~\cite{mariantoni2011implementing}. Ensembles of quantum systems are natural platforms for QMs, given their large storage capacity. Multiple qubits can be stored taking advantage of direct spatial addressing to access different regions of the ensemble~\cite{jiang2019experimental, mariantoni2011implementing}. However, for ensembles in the solid state that offer prospects for high-density QMs and typically have inhomogenous broadening, \emph{spectral}-addressing can be used to distinguish excitations stored collectively in the ensemble. 
Solid-state atomic ensembles have long coherence times for both microwave~\cite{wolfowicz2013atomic, ranjan2020multimode, steger2012quantum,ortu2018simultaneous, bar2013solid} and optical~\cite{ortu2018simultaneous, bar2013solid, sukachev2017silicon} transitions and can couple to resonant cavities facilitating read, write and control operations. 
Coherent control allows coherence times to be extended by dynamical decoupling (DD)~\cite{bar2013solid, naydenov2011dynamical, carr1954effects, meiboom1958modified} or transferring the qubit state to a more coherent transition~\cite{morton2008solid}. Coupling the ensemble to a cavity in the strong coupling regime~\cite{kubo2010strong, weichselbaumer2020echo, probst2013anisotropic,schuster2010high, zhong2017interfacing} or with cooperativity $C=1$ facilitates writing and reading information with unit efficiency~\cite{afzelius2013proposal, julsgaard2013quantum}.% ($C\equiv g_{\rm ens}^2/\kappa\Gamma$, where $g_{\rm ens}$ is the ensemble coupling rate while $\kappa$ and $\Gamma$ are the linewidths of the cavity and ensemble, respectively).

% One of the simplest memory protocols in inhomogeneously broadened systems is 
One of the simplest memory protocols in inhomogeneously broadended systems is the Hahn echo~\cite{hahn1950spin}, in which an excitation stored within an ensemble is inverted by a single $\pi$ pulse and re-emitted later as an `echo'. 
This has been used  widely for retrieval of weak  excitations in multimode microwave~\cite{probst2015microwave, ranjan2020multimode, grezes2014multimode} and optical~\cite{lovric2013faithful, lvovsky2009optical} memories and indeed formed the basis of early (classical) information storage proposals~\cite{anderson1955spin}. However, the simple Hahn echo approach is unsuitable for quantum memories, as it leads to amplified (and thus noisy) emission from the quantum systems in their excited state~\cite{ruggiero2009two}.
One solution is using two $\pi$ pulses to return the ensemble predominantly to the ground state before the memory is accessed~\cite{damon2011revival, julsgaard2013quantum}. This requires suppressing the emission of the echo that would appear after the first $\pi$ pulse. A second limitation of the Hahn echo sequence is that it acts as a `first-in last-out' memory, rather than permitting random access to stored qubits. 
Various approaches have been explored to address these limitations: frequency-tunable cavities can be shifted off-resonance except when the desired echo is being emitted~\cite{kubo2011hybrid}; 
external electric or magnetic field gradients can be used to label stored excitations~\cite{kraus2006quantum}; 
and AC Stark shifts can shift the emission of excitations into different time bins~\cite{zhong2017nanophotonic}. None of these ingredients alone realises a random access QM and the requirement of additional control fields poses a significant practical limitation.
%For example, Ref \cite{wu2010storage} demonstrated the retrieval of two microwave excitations from a spin ensemble in either order, however the applied magnetic field gradients have poor compatibility with the superconducting resonators used to achieve high-cooperativity coupling and the superconducting qubits which may be interfaced to such memories.
For example, the magnetic field gradients used in Ref.~\cite{wu2010storage} to retrieve two microwave excitations in arbitrary order have poor compatibility with the superconducting resonators and qubits. % used to achieve high-cooperativity and provide the quantum states to be stored respectively.
In this Article we introduce a simpler and more powerful approach to labelling and recalling stored qubits from an ensemble QM, using chirped control pulses. This protocol suppresses unwanted echoes, allows random access to multiple memory modes (read and write operations performed in arbitrary order), and naturally embeds DD. We demonstrate the performance of the protocol in a prototype memory using weak microwave excitations stored within donor spins in silicon.

\begin{figure}
    \centering
    \includegraphics[width=0.48\textwidth]{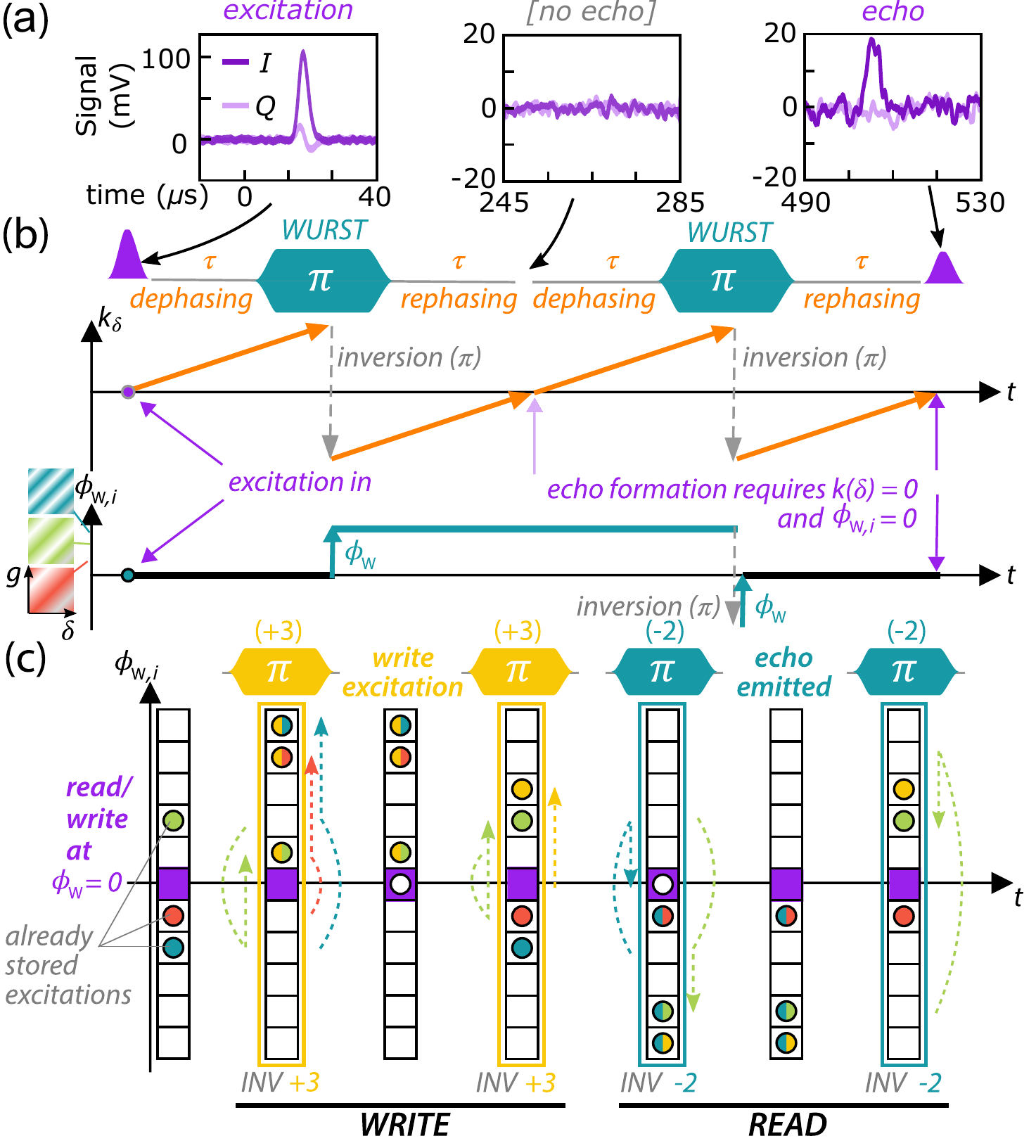}
    \caption{
    {\bf Using chirped pulses to silence echoes and encode quantum information.}
    (a) A weak ($\langle n \rangle\sim 200$ photon) Gaussian microwave excitation is applied to an ensemble of Bi donor spins in silicon followed by two chirped `WURST' $\pi$-pulses (chirp rate 20~MHz/ms 200~$\upmu$s duration). A spin echo is observed only after the second WURST pulse. The demodulated I and Q quadratures are shown.
    (b) The WURST pulse (index $i$) inverts the phase of the ensemble and imprints a phase pattern $\phi_{\mathrm{W},i}$ in the space of the atom-cavity coupling $g_0$, and the frequency detuning $\delta$.
    $k_{\delta}$ describes a wavevector of phase acquisition in frequency-space, which grows linearly in time by the precession of the inhomogenously broadened ensemble. Excitations are stored in the ensemble at $k_{\delta}=\phi_{\rm W}=0$. 
    A WURST pulse is applied at time $\tau$, and a further $\tau$ after the pulse the spin-wave due to inhomogeneous broadening is refocused ($k_{\delta}=0$), however, the WURST-imprinted phase pattern $\phi_{\rm W}$ remains, suppressing echo emission.
    A further period of $\tau$-WURST-$\tau$ returns the excitation to $\phi_{\rm W}(g,\delta)=0$ and an echo is emitted. 
    (c) The function $\phi_{\mathrm{W},i}(g,\delta)$ is defined by the chirped pulse parameters, and can be used as a storage index to encode quantum excitations into the quantum memory. We illustrate WRITE and READ operations: Applying a WURST pulse (illustrative $\phi_{\mathrm{W},i}$ index: +3) before and after some excitation, it becomes stored within the corresponding region in $\phi_{\rm W}$-space leaving previously written excitations unaffected. The same procedure addressing a previously stored excitation causes it to be emitted as an echo. }
    \label{fig:1}
\end{figure}

Sweeping the frequency of a control pulse across that of an atomic transition can be used to realise an inversion (or $\pi$ pulse) by `adiabatic fast passage' (AFP)~\cite{Baum1985,Kupce1995,Kupce1995a,Garwood2001}. Such chirped pulses, with forms such as `wideband, uniform rate, smooth truncation' (WURST)~\cite{o2013wurst} (see \SI~for further details) among others e.g.~\cite{malinovsky2001general, tesiram2010implementation} --- offer several important advantages for control of ensemble QMs. 
First, they offer robustness to variations in the strength of the control field (e.g.\ laser field or microwave magnetic field) --- particularly attractive for planar microresonators with inhomogeneous coupling~\cite{sigillito2014fast}. 
Second, the wide bandwidth pulses refocus most spins within the cavity linewidth.
Third, when used to refocus an inhomogenously broadended ensemble, a \emph{pair} of pulses must be applied to produce an echo -- inherently suppressing the first echo as demonstrated in the optical domain~\cite{gerasimov2017quantum, damon2011revival, bonarota2014photon}, 
and shown in Fig.~\ref{fig:1}(a). 
The final feature, which we explore further below, is that the chirped pulse can be used to imprint a pulse-specific phase distribution across the ensemble, allowing multiple excitations to be independently stored and retrieved within the ensemble QM.

Our random-access QM protocol assumes an inhomogeneously broadened ensemble of two-level systems (e.g.\ spins) coupled to a cavity, which can be decomposed into sub-ensembles with transition frequency detuning $\delta$ and spin-cavity coupling $g_0$. After a state has been stored collectively in the ensemble, precession of the sub-ensembles over some time $\tau$ (the `dephasing' period) causes the excitation to evolve into a spin wave with wavevector $k_{\delta}$ in spin-detuning (Fig.~\ref{fig:1}(b)). A WURST `$\pi$-pulse' inverts the phase acquired by the spins, and imparts some additional phase shift $\phi_{\rm W}$, that varies between sub-ensembles, and is a function of WURST parameters (chirp rate and pulse amplitude). Following a subsequent `rephasing' period of $\tau$, the spin wave defined by $k_{\delta}$ is refocused, however, due to the phase pattern defined by $\phi_{\rm W}$, there is no ensemble coherence and no collective emission of an echo. A second \emph{identical} WURST pulse unwinds the phase pattern $\phi_{\rm W}$, so that when $k_{\delta}$ next returns to 0, the initial excitation is emitted as an echo. 
We exploit this behaviour to achieve a general random-access QM: the phase patterns $\phi_{{\rm W},i}$, defined by different WURST pulses, provide the storage index for the memory, as shown in Fig.~\ref{fig:1}(c). The application of a WURST pulse before and after an excitation, `writes' it into the ensemble without affecting previously stored excitations. The same pair of WURST pulses is used later to read out the excitation, enabling random access of the QM. Applying WURST pulses in identical pairs ensures that stored excitations are unaffected by any read/write operation, beyond introducing  periodic inversions that offer built-in dynamical decoupling. 

In summary, the random access protocol consists of applying repeating blocks of the form $(\pi_j - \msquare - \pi_j)$ where $\pi_j$ is a unique WURST $\pi$-pulse addressing a particular storage index $j$. $\msquare$ can be i) a weak input excitation to be stored in location $j$ of the memory; ii) an echo, constituting an excitation being retrieved from location $j$ in memory; or iii) null, for a clock cycle in which no information is being read or written, where the block is merely applied for dynamical decoupling with $\pi_j$ addressing an unused memory mode.

\begin{figure}
    \centering
    \includegraphics[width=0.48\textwidth]{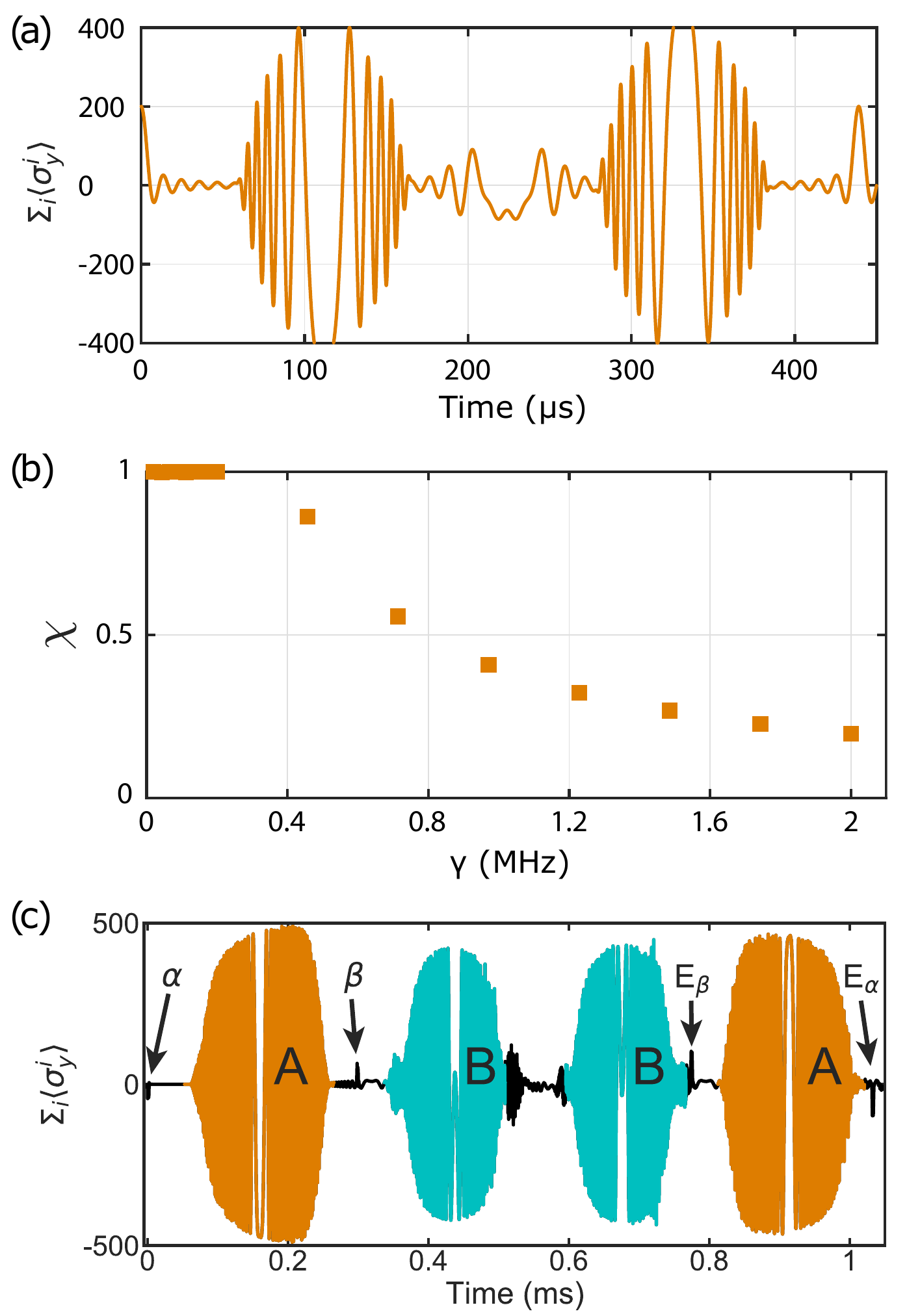}
    \caption{{\bf Simulation of the memory protocol.}
    Numerical simulations of 3.23$\times10^5$ spins with 100~kHz linewidth showing the effects of a WURST pulse on a large ensemble of emitters. (a) Expectation values $\langle\hat{\sigma}^i_y\rangle$ summed over all spins shows that, following a pair of identical WURST pulses, an initial excitation is faithfully recovered as an echo.
  (b) The ratio of the excitation to echo amplitude ($\chi$) is shown as a function of the spin linewidth $\gamma$ for a WURST pulse of 0.5~MHz bandwidth. For spin linewidths within the WURST bandwidth the sequence has approximately unit efficiency. (c) Retrieval of two weak excitations within an `ABBA' sequence composed of two different  WURST pulses (A and B) with different chirp rates, $R_A= -2\pi\times 11.25 $ MHz/ms and $R_B= 2\pi \times 7.50$ MHz/ms. Two weak pulses, $\alpha$ and $\beta$, are stored in the spins and re-emitted at the expected tims as echoes $E_\alpha$ and $E_\beta$. %The pulses explore the same finite frequency interval and have the same amplitude.
  }
    \label{fig:theory_efficiency}
\end{figure}

We next validate the theoretical performance of the protocol through its constituent elements to confirm it can support high fidelity storage. 
We numerically calculate the spin dynamics by discretizing an inhomogeneously broadened spin ensemble into $10^5$ subensembles and solve their individual dynamics subject to chirped pulses. 
In Fig.~\ref{fig:theory_efficiency}(a,b) we study a pair of identical WURST pulses, plotting spin expectation values $\langle\hat{\sigma}^i_y\rangle$ summed over all spins in an ensemble when the WURST pulse chirps $\pm0.25$~MHz in 100$\upmu$s showing that the initial peak value is well recovered.  
As long as the spin linewidth, $\gamma \lesssim \Delta_f$, the bandwidth of the WURST pulse, the refocusing effect yields approximately unit efficiency. This supports the use of repeated WURST pulses in our protocol and verifies that these will not limit our memory efficiency, at least for a core of sufficiently coupled spins. Next in Fig.~\ref{fig:theory_efficiency}~(c) we examine the simplest sequence illustrating random access in the memory protocol, where two distinct WURST pulses are used (`A') and (`B'), with chirp rates of $-2\pi\times 11.25 $ MHz/ms and $2\pi \times 7.50$ MHz/ms respectively. The simulation considers two input excitations, $\alpha$ and $\beta$, applied in the sequence $\alpha - A - \beta - B - B - E_\beta- A - E_\alpha$. We see the input $\beta$ is remitted as echo $E_\beta$ following the second $B$ pulse, while $\alpha$ is remitted as $E_\alpha$ only after the second $A$ pulse, confirming the basic principle of the random access protocol.

To further validate the protocol we perform experiments on a prototype device. The memory consists of an ensemble of bismuth donor spins in natural silicon, coupled to a planar superconducting niobium resonator at 100~mK with resonant frequency 7.093~GHz and quality factor $\sim$18,000.  The bismuth donors are implanted at a target density of $10^{17}$~cm$^{-3}$ in the top 1~$\upmu$m of the sample.  Further details of the chip and setup are provided in the supplementary information. A schematic of the device, its magnetic field and the implantation profile is shown in in (\SI) Fig.~S1, a schematic of the full setup is shown in (\SI) Fig.~S2 and details of resonator performance are provided in (\SI) Fig.~S3. 

Fitting to S-parameter measurements of the cavity gives a cooperativity $C = 0.07(2)$ (see SI \cite{abe2011electron}), leading to a predicted one-way memory efficiency of $\eta_{\rm em} = \frac{4C}{(1+C)^2}$ of 0.2 (Ref~\cite{afzelius2013proposal})). Although this is larger than previously reported values for ensemble microwave memories of 0.01--0.04~\cite{ranjan2020multimode,grezes2014multimode,probst2015microwave}, it poses a bound for the amplitude of the retrieved excitations in our demonstration of the random access QM protocol below.

\begin{figure}[t]
    \centering
    \includegraphics[width=0.48\textwidth]{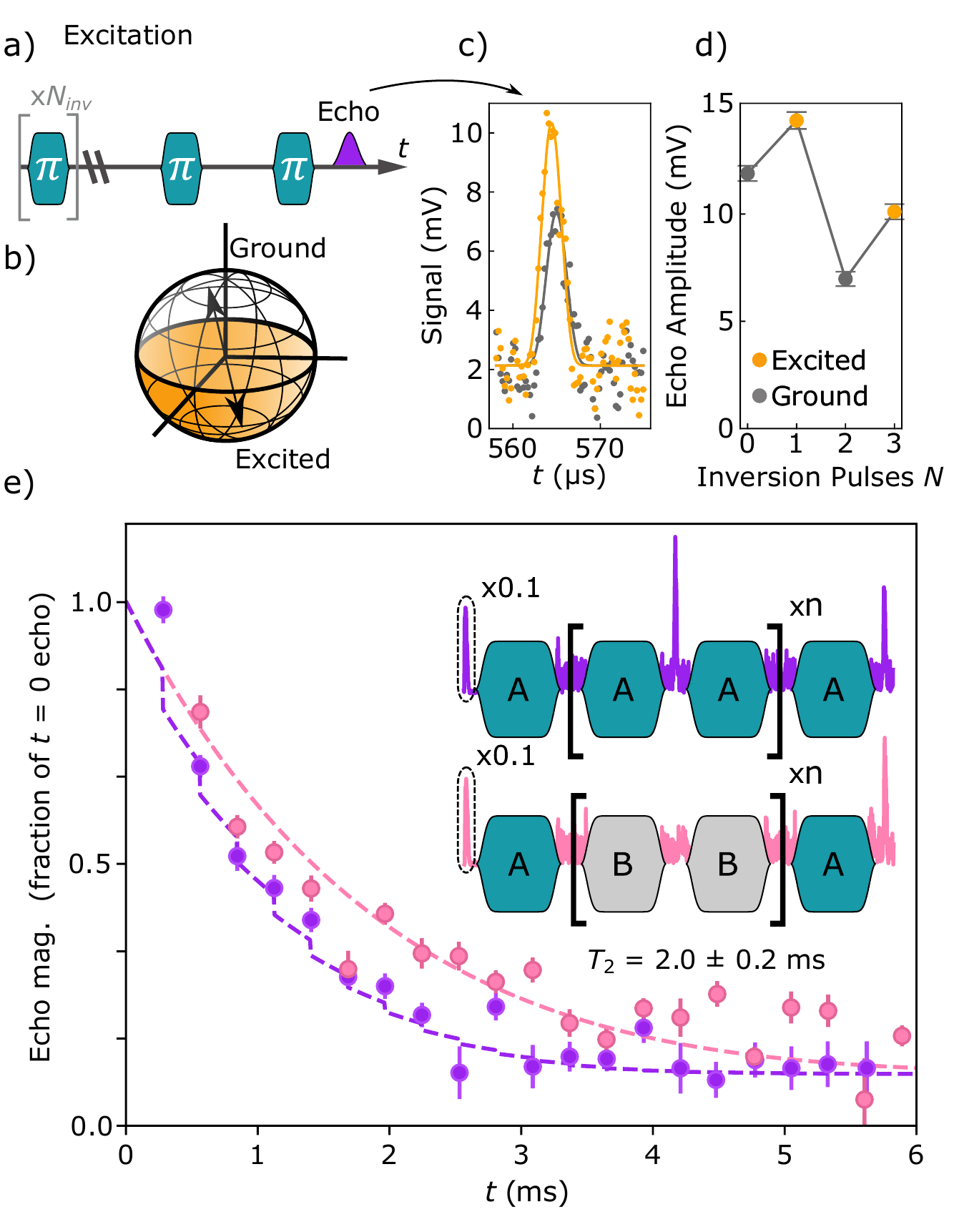}
    \caption{
    {\bf Demonstrating the importance of suppressed echo emission.}
    (a) The pulse sequence used to study the echo following a pair of WURST pulses as a function of the state of the ensemble. 
    (b) The approximate ensemble ground and excited state in a Bloch sphere representation of the ensemble magnetisation.  
    (c) Two example echoes emitted from the ground state ($N_{\rm inv} = 2$) or inverted state ($N_{\rm inv} = 3$). Larger amplitude echoes are seen from the inverted ensembles due to amplification of the echo, which adds noise to a quantum memory. 
    (d) Echo amplitude as a function of $N_{\rm inv}$. The overall reduction of echo amplitude with increasing $N_{\rm inv}$ is likely due to gradual saturation of the spin ensemble. 
    (e) Coherence decay rates from different WURST dynamical decoupling sequences. Given identical pulses (A[AA]$_n$A - purple) echo emission occurs every second pulse leading to an accelerated decay in coherence compared to an A[BB]$_n$A (pink) sequence where only one echo is emitted, at the end. 
    A fit to the A[BB]$_n$A data (dashed line) gives a coherence time of 2~ms, while fitting to the A[AA]$_n$A data accounting for additional loss from repeated echoes (dashed line, see  Eq.~5 \SI) gives a memory efficiency $\eta_{\rm em}$ of 0.17. Error bars represent one standard deviation from fitting Gaussians to echoes.}
    \label{fig:2}
\end{figure}

Despite the spin-cavity coupling in these experiments being well below unit cooperativity, we can already see the importance of suppressing echo emission from an inverted ensemble. 
In Fig.~\ref{fig:2}(a-d) we apply $N_{\rm inv}$ WURST $\pi$-pulses (for $N_{\rm inv}$=0--3) before an echo sequence to selectively prepare the ensemble into a ground ($N_{\rm inv}$ even) or inverted ($N_{\rm inv}$ odd) state. Using weak excitations of $\langle n \rangle \sim 200$ microwave photons ensures that the ensemble is only weakly perturbed from the ground or excited state when emitting. % 
We see that echoes emitted from an inverted ensemble have larger amplitude than those emitted from a (quasi-)ground state ensemble, indicating that there has been amplification of the input signal from stimulated emission (the echo is always weaker than the input signal due to other losses). This amplification of states as they are emitted is incompatible with high fidelity state retrieval, and is why robust quantum memory schemes must ensure emission from a (quasi-)ground state ensemble, as supported by our protocol .
%\textcolor{red}{is this no cloning? if so lets say so here.} 

By increasing the delay between WURST pulses, we measure a coherence time of this prototype memory, $T_2 \sim 0.6~$ms (Fig.~S3). This lifetime can be extended using dynamical decoupling (DD) sequences which involve a train of repeated $\pi$-pulses~\cite{meiboom1958modified,viola1999dynamical}.
If the DD sequence uses identical WURST pulses (i.e. A[AA]$_n$A), an echo is emitted after every pair of pulses as shown in the upper inset to Fig.\ref{fig:2}(e). This repeated echo emission can be avoided by introducing a second type of WURST pulse (`B') of different chirp rate and/or amplitude to the first (`A'), creating the sequence A[BB]$_n$A, shown in the lower inset to Fig.\ref{fig:2}(e).
Only one echo appears in this sequence, after the second `A' pulse which is applied at the end, and the decay time constant of this echo provides a measure of the memory storage lifetime, $T_{\rm M}= 2.0(2)$~ms. This lifetime is about three times longer than that measured with a Hahn echo, thanks to the effect of DD. 
The A[AA]$_n$A WURST DD sequence --- in which echoes are emitted periodically throughout --- gives a decay time constant \emph{shorter} than $T_{\rm M}$. We model this assuming a constant fraction $\eta_{\rm em}$ of the excitation is lost with each echo, in addition to exponential decay with time constant $T_{\rm M}$. Fitting this model yields $\eta_{\rm em}=0.17(7)$, and interpreting this value as the one-way efficiency we find good agreement with the value extracted from S-parameter measurements.
In a high efficiency quantum memory ($C\sim\eta_{\rm em}\sim1$), DD sequences of this type will be essential to avoid premature emission of stored excitations.

\begin{figure}
    \centering
    \includegraphics[width=0.48\textwidth]{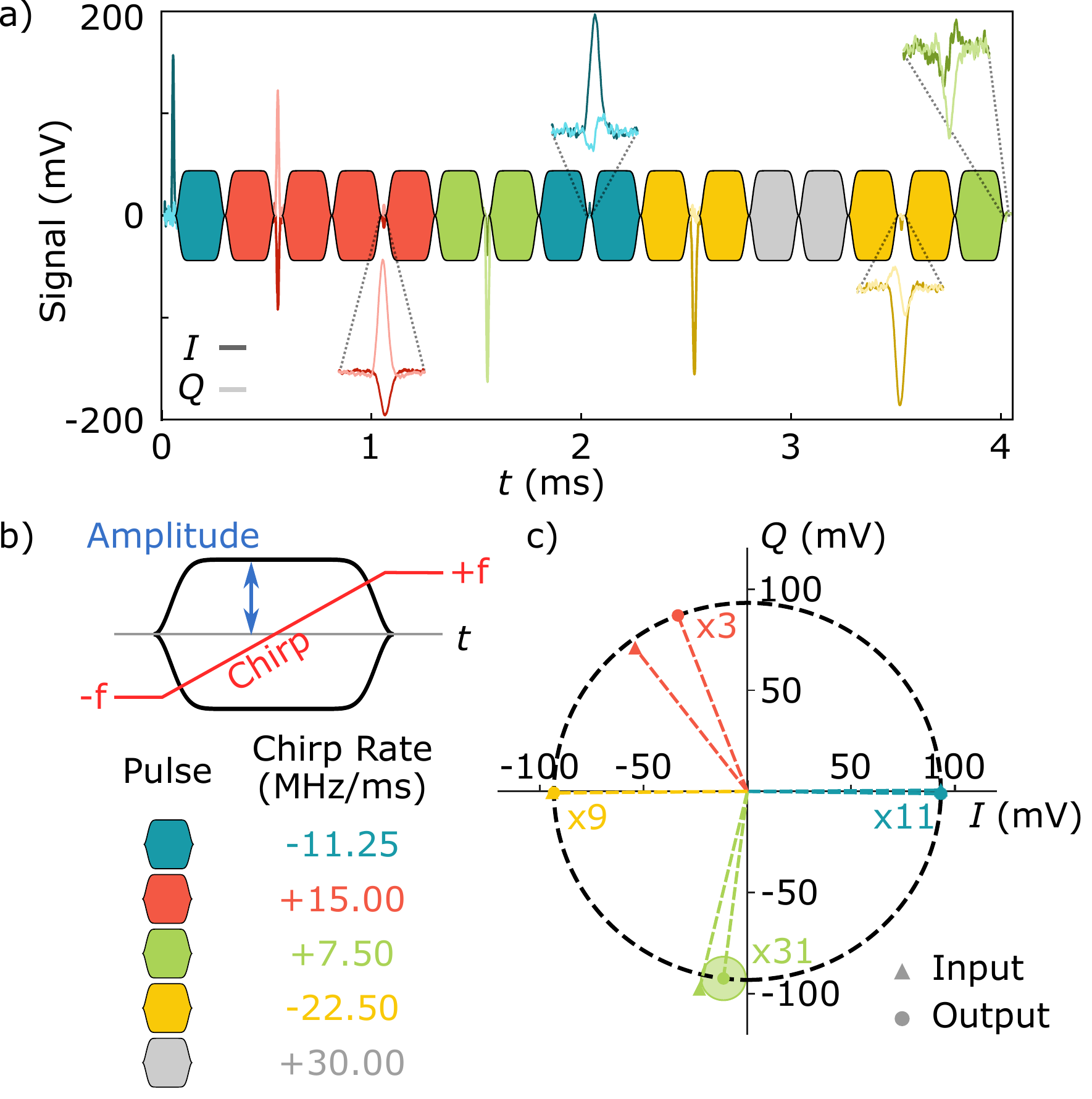}
    \caption{
    {\bf Experimental demonstration of the random access memory protocol.}
    (a) An illustrative segment of the protocol in which WURST pulses, excitations, and echoes are colour coded, with WURST parameters as defined in (b).
    In total four excitations are stored for times up to 2~ms. Excitations and echoes are shown with magnified echoes offset from the main trace. The first and last pulses do not come in pairs as the memory sequence starts and finishes half way through a clock cycle.
    %(b) The parameters of the WURST pulses used in (a). 
    (c) Echoes and excitations on an Argand diagram matching their phases. The magnitude of each echo is rescaled to account for losses from the finite spin-resonator cooperativity in this experiment and spin decoherence.
    %
    %NOTE: This figure is intended to be displayed at two-column width
    }
    \label{fig:3}
\end{figure}

Having confirmed control of echo emission using chirped pulses, we present in Figure~\ref{fig:3}~(a) an experimental demonstration of the random access protocol introduced earlier, showing the storage, protection and retrieval of four weak ($\langle n
\rangle \sim 1200$ photon) microwave excitations, using five distinct WURST pulses (see Fig.~\ref{fig:3}~(b)). The photon number was calibrated from measurements of Rabi frequency and Purcell relaxation (see SI and Fig.~S2).
%
%\onecolumngrid
%
%\twocolumngrid
%
Each excitation is encoded into the memory using a pair of identical WURST pulses (colour-coded in teal, coral, lime and mustard), and retrieved later (storage time varying from 0.5 to 2.5~ms) by applying the same pair of pulses. Each echo can be unambiguously matched to one of the input excitations through its phase (see Fig.~\ref{fig:3}~(c)), which we confirm by repeating permutations of the sequence with only one excitation present (Fig.~S10). A fifth variant of WURST pulse (shown in grey) is used to perform DD. The weak amplitude of retrieved signals relative to input states is due primarily to the limited cooperativity in these experiments, rather than the control fidelity in the random-access protocol or decoherence (see \SI~\S D). Rescaling the echo amplitudes by the factors shown in Fig.~\ref{fig:3}~(c) we observe that the phase of the excitation is generally well preserved.
The largest phase error, from the second excitation (coral) is attributed to a phase shift from the Josephson parametric amplifier, as the echo is of larger amplitude due to the short storage time (Fig.~S4).

\begin{figure}
    \centering
    \includegraphics[width=0.48\textwidth]{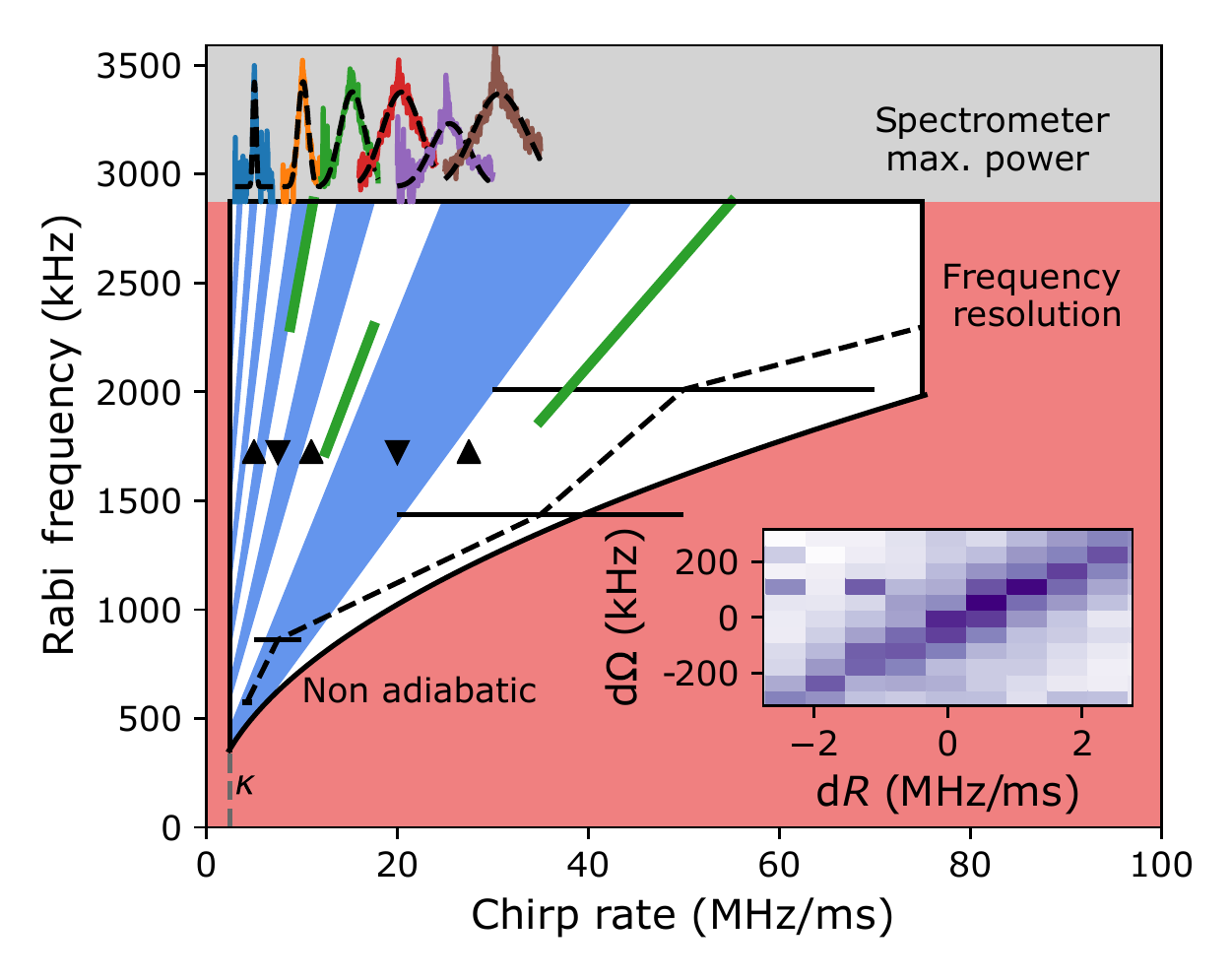}
    \caption{
    \textbf{Counting memory modes.}
    Each memory mode is determined by the phase pattern $\phi_{\rm W}$ imparted by a WURST pulse.
    The parameter space where WURST pulses efficiently refocus spins is bounded in black. Bounds are the cavity linewidth, the maximum spectrometer power and requirements the WURST is adiabatic. A frequency resolution limit occurs due to the homodyne scheme and would be removed in a heterodyne scheme. We measure the adiabaticity limit by thresholding echo intensity shown in the \SI~and here by the black dashed line giving good agreement with the theoretical line. 
    We measure distinctiveness of WURST pulses using AB-echo sequences. The results of sweeping the B pulse chirp rate at maximum power are inset to the top of the figure and fit by Gaussian profiles. We show a two dimensional map sweeping chirp rate and Rabi frequency in the inset. In 2D sweeps there are lines of WURST pulses which impart the same $\phi_{\rm W}$ function to the spin ensemble. Green lines in the main panel show the line of equivalent WURST pulses from other two dimensional maps. 
    Interpolating data from AB-echo sweeps we determine the width of a line of equivalent WURST pulses at fixed Rabi frequency. We partition the space of available WURSTs into unique WURSTs (i.e. memory modes) by constraining neighbouring WURST pulses to be separated by the half width hundredth max and show an example partitioning where blue/white sections indicate unique WURST pulses. 
    We show the WURST pulses used in Fig.~\ref{fig:3} where the up/down markers refer to positive/negative chirp direction.}
    \label{fig:fig4}
\end{figure}

The memory capacity is determined by the number of independent WURST pulses that can be used in the protocol. This is related to the spectral width ($\Delta_f$) of the storage ensemble, the narrower of the cavity linewidth $\kappa$ and that of the inhomogeneous ensemble. 
The WURST pulses can be parameterised using their chirp rate $R$ and amplitude $A_{\rm W}$%($A_{\rm W} = 1$ gives Rabi frequency $\Omega = 2.87$~MHz)
, bounds illustrated in Fig.~\ref{fig:fig4}. Adiabiticity of refocusing imposes a lower bound of $A_{\rm W}$~\cite{doll2013adiabatic}, with an upper bound set by the maximum pulse amplitude, limited for example by the pulse amplifier or sample heating). We determined the adiabatic bounds on $A_{\rm W}$ in our experiment through microwave simulations, confirmed through experiments of the type illustrated in Fig.~\ref{fig:1}(a) (Fig.~S6). The rate $R$ has some upper bound set by the experimental frequency resolution and a lower bound determined by the need to at least chirp across $\Delta_f$ within the effective duration of the WURST pulse ($T_{\rm W, eff}$), such that $R\gg\Delta_f/T_{\rm W, eff}$. 

The separation of distinct pulses in $R$ and $A_{\rm W}$ is governed by the requirement of independence in the storage modes of the memory. We explore this using pulse sequences of the form $\alpha - \pi_{\rm A} - \pi_{\rm B} - {\rm [echo]}$, where an excitation $\alpha$ is followed by two WURST pulses whose parameters are varied. For sufficiently distinct $\pi_{\rm A}$ and $\pi_{\rm B}$, no echo should be observed.
First, we fix the parameters of $\pi_{\rm A}$ and vary those of $\pi_{\rm B}$, shown in the inset to Fig.~\ref{fig:fig4}. Equivalent WURST pulses lie on lines of positive gradient --- as $R$ increases, the phase from the WURST pulse decreases, compensated by increasing $A_{\rm W}$. Additional two-dimensional parameter sweeps are shown as green lines in Fig.~\ref{fig:fig4}, we also acquire several one-dimensional sweeps (a subset are shown in Fig.~\ref{fig:fig4}). We extract the different regions of WURST pulses equivalence requiring that each WURST pulse is separated by at least the half width, \emph{hundredth} max of the neighbouring regions (Fig.~S7). We find $\sim$8 distinct WURST pulses at an amplitude of 0.6~V, and the same from chirping in the reverse direction, giving $\sim$16 distinct memory modes.

\begin{figure}
    \centering
    \includegraphics[width=0.48\textwidth]{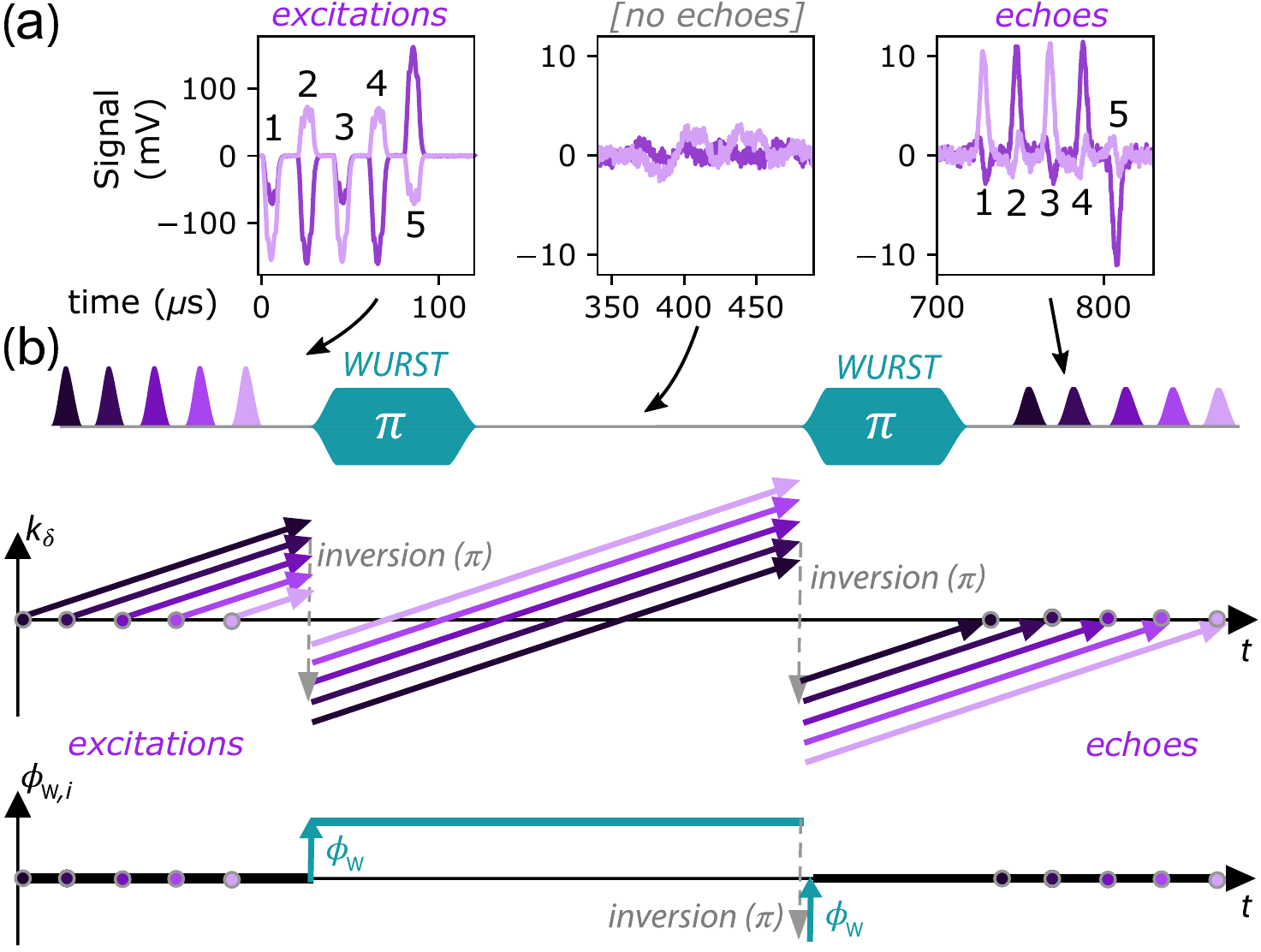}
    \caption{\textbf{First in first out (FIFO) time bin encoding of multiple echoes.} (a) An experimental demonstration of multiple excitations stored using the FIFO protocol with two identical WURST pulses. The pattern of excitation/echo phases allows echoes and excitations to be unambiguously matched.  (b) Schematic showing evolution of spin waves and WURST phase resulting in the FIFO encoding with silenced echoes.}
    \label{fig:fifo}
\end{figure}

Chirped pulse encoding can be combined with other methods such as time-bin encoding (used to store up to 100 weak microwave excitations~\cite{wu2010storage,ranjan2020multimode}) to implement a memory offering random access to large registers. For example, in Fig.~\ref{fig:fifo} we demonstrate how a pair of WURST pulses can be used to store a register of five microwave excitations, later retrieved in the same order in which they were written (a first-in first-out, or FIFO memory). Therefore, by replacing the single excitations shown in Fig.~\ref{fig:3} with such registers, the capacity of the quantum memory is further extended. 
Additional strategies for increasing the storage capacity include addressing the instrumentation-limited bounds on $R$ and $A_{\rm W}$, and by increasing the WURST pulse duration (at the expense of slower read/write speeds and less effective DD).

The memory efficiency in the experimental demonstration was primarily limited by the spin-resonator cooperativity ($C\sim0.06$), and there are several practical approaches being pursuedto increase this to $C\sim1$ using Bi donors and superconducting microresonator cavities by increasing the spin-cavity coupling $g_0$ and the number of resonant spins (see Ref~\cite{o2020spin}).
%
%Other factors limiting efficiency include finite coherence time and imperfect pulses. 
Although the memory lifetime in this demonstration using a natural-silicon host was only 2~ms, rare-earth spins coupled to similar resonators have been shown to have coherence times of tens of milliseconds~\cite{Dantec2021} while coherence times over 300~ms have been shown for near-surface Bi donors in isotopically enriched $^{28}$Si~\cite{ranjan2020multimode}.

The chirped pulse protocol introduced here could also be applied to provide random read/write access in optical QMs. 
Chirped microwave pulses could be applied to directly drive transitions in spin-active optical QMs, such as Nd in YVO$_4$~\cite{zhong2017nanophotonic}, providing DD and controlling access to a register of stored optical excitations.
Furthermore, adiabatic fast passages (achieved using acousto-optic and electro-optic modulators) have been used to optimally invert optical transitions in demonstrations of optical QMs~\cite{damon2011revival}. Similarly chirped optical pulses of varying parameters could allow for a random access protocol to be directly implemented in the optical domain, talking advantage of the stronger atom-cavity coupling and larger cavity bandwidth.

\section{Acknowledgements}
We thank Philippe Goldner for insightful discussions relating to the implementation of this protocol in the optical domain. We thank the UK National Ion Beam Centre (UKNIBC) where the silicon samples were ion implanted and Nianhua Peng who performed the ion implantation. This work has received funding from the U.K. Engineering and Physical Sciences Research Council (EPSRC), through UCLQ postdoctoral fellowships (O.W.K, M.S.) Grant No.\ EP/P510270/1 and a Doctoral Training Grant (J.O'S.). J.J.L.M. acknowledges funding from the European Research Council under the European Union's Horizon 2020 research and innovation programme (Grant agreement No.\ 771493 (LOQO-MOTIONS).

\section{Author Contributions}
J.O'S and O.W.K. performed the experiments with assistance from C.W.Z.. J.O'S, O.W.K. and J.J.L.M. analysed the data with input from J.A. and M.S..  J.A. and J.J.L.M designed the random access protocol with input from O.W.K. and J.O'S. K.D. and K.M. performed numerical analyses and the analytical treatment. C.T. and S.W. fabricated the device. A.H. and I.S. assembled, tested, and provided scientific support in the operation of the JPA. J.O'S, O.W.K. and J.J.L.M. wrote the manuscript with input from all authors. 

\section{Data Availability}
The datasets generated and analysed during the current study are available in the UCL Research Data repository, doi.org/10.5522/04/14541747.

\bibliography{bibliography}

\newpage

\end{document}

\beginsupplement

%\begin{abstract}
%Superconducting qubits Quantum memories are a critical part of many 
%\end{abstract}

\title{Supplementary Information for Random-access quantum memory using chirped pulse phase encoding}

\author{James~O'Sullivan}
\altaffiliation{These authors have contributed equally to this work} 
\affiliation{London Centre for Nanotechnology, UCL, 17-19 Gordon Street, London, WC1H 0AH, UK}
\author{Oscar~W.~Kennedy}
\altaffiliation{These authors have contributed equally to this work} 
\affiliation{London Centre for Nanotechnology, UCL, 17-19 Gordon Street, London, WC1H 0AH, UK}
\author{Kamanasish Debnath}
\affiliation{Department of Physics and Astronomy, Aarhus University, DK-8000 Aarhus C, Denmark}
\author{Joseph Alexander}
\affiliation{London Centre for Nanotechnology, UCL, 17-19 Gordon Street, London, WC1H 0AH, UK}
\author{Christoph~W.~Zollitsch}
\affiliation{London Centre for Nanotechnology, UCL, 17-19 Gordon Street, London, WC1H 0AH, UK}
\author{Mantas~\v{S}im\.{e}nas}
\affiliation{London Centre for Nanotechnology, UCL, 17-19 Gordon Street, London, WC1H 0AH, UK}
\author{Akel Hashim}
\affiliation{Lawrence Berkeley National Laboratory, Berkeley, CA 94720, USA}
\author{Christopher~N.~Thomas}
\affiliation{Cavendish Laboratory, University of Cambridge, JJ Thomson Ave,  Cambridge CB3 0HE, UK}
\author{Stafford Withington}
\affiliation{Cavendish Laboratory, University of Cambridge, JJ Thomson Ave,  Cambridge CB3 0HE, UK}
\author{Irfan Siddiqi}
\affiliation{Lawrence Berkeley National Laboratory, Berkeley, CA 94720, USA}
\author{Klaus M\o{}lmer}
\affiliation{Department of Physics and Astronomy, Aarhus University, DK-8000 Aarhus C, Denmark}
\author{John~J.~L.~Morton}
\affiliation{London Centre for Nanotechnology, UCL, 17-19 Gordon Street, London, WC1H 0AH, UK}
\affiliation{Department of Electrical and Electronic Engineering, UCL, Malet Place, London, WC1E 7JE, UK}

\maketitle

\section{Methods}
\subsection{Device fabrication}
\begin{figure}
    \centering
    \includegraphics[width=0.48\textwidth]{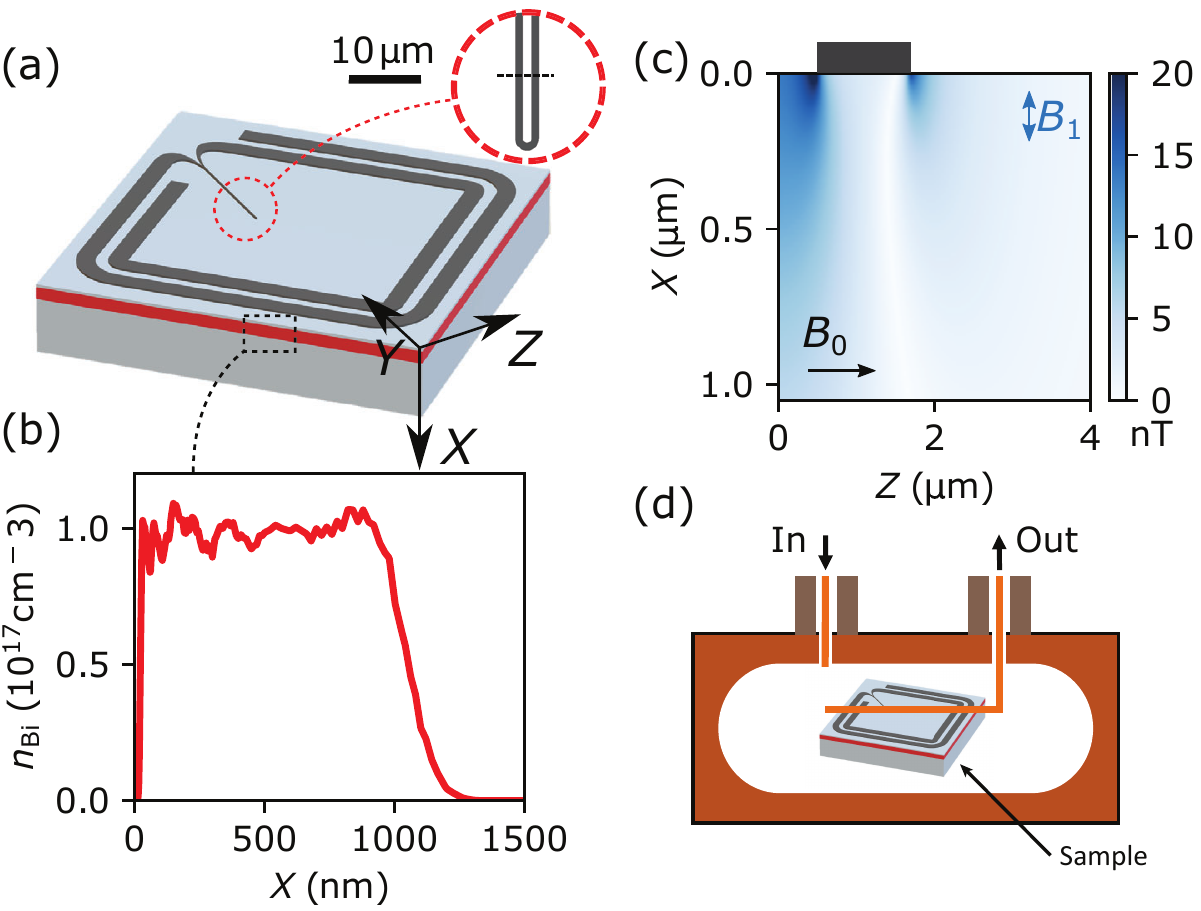}
    \caption{
    {\bf The prototype memory.}
    (a) Illustration of the silicon wafer with near-surface implanted Bi layer and Nb resonator patterned on the surface.
    (b) Simulated Bi implantation profile as a function of depth below the substrate surface.
    (c) Finite element simulations of the magnetic field generated in the vicinity of the resonator inductor
    (d) Illustration of the copper sample box with asymmetric antennae.
    }
    \label{fig:sample_schematic}
\end{figure}

A float-zone silicon wafer of natural isotopic abundance is ion implanted with bismuth at a chain of energies targeting a density of $10^{17}$~cm$^{-3}$ bismuth in the top 1~$\upmu$m of the sample. This is annealed at 900$^\circ$C for 5 minutes to incorporate the bismuth into the silicon matrix forming spin-active donors.
A planar niobium microresonator is patterned on the top of the sample by liftoff. The Nb film is 100~nm thick and has a field dependent frequency (see Ref. \cite{o2020spin}) of 7.093~GHz at a magnetic field of 46~mT. The resonator is a lumped element design comprising a pair of parallel capacitive plates and a narrow doubled-back inductor wire, designed to enhance the magnetic field generated by the resonator close to the substrate surface. The sample is placed inside a copper cavity, as shown in Fig.~\ref{fig:sample_schematic}(d). Two antennae extend into the box; one short stub antenna (high insertion loss $\gtrsim 30$~dB, low coupling) is connected to a 30~dB attenuated microwave in-line and used to apply microwave excitations. Another long antenna (large coupling) is used to collect microwave signals and is connected to the amplification chain. The asymmetry results in a large collection efficiency of spin-echo signals.

A schematic of the device is shown in Fig.~\ref{fig:sample_schematic}(a), the simulated bismuth implantation profile in Fig.~\ref{fig:sample_schematic}(b), the simulated microwave magnetic fields caused by the zero point fluctuations in the resonator in Fig.~\ref{fig:sample_schematic}(c) and a schematic of the 3D cavity in Fig.~\ref{fig:sample_schematic}(d).

\begin{figure}
	\includegraphics[width = 0.5\textwidth]{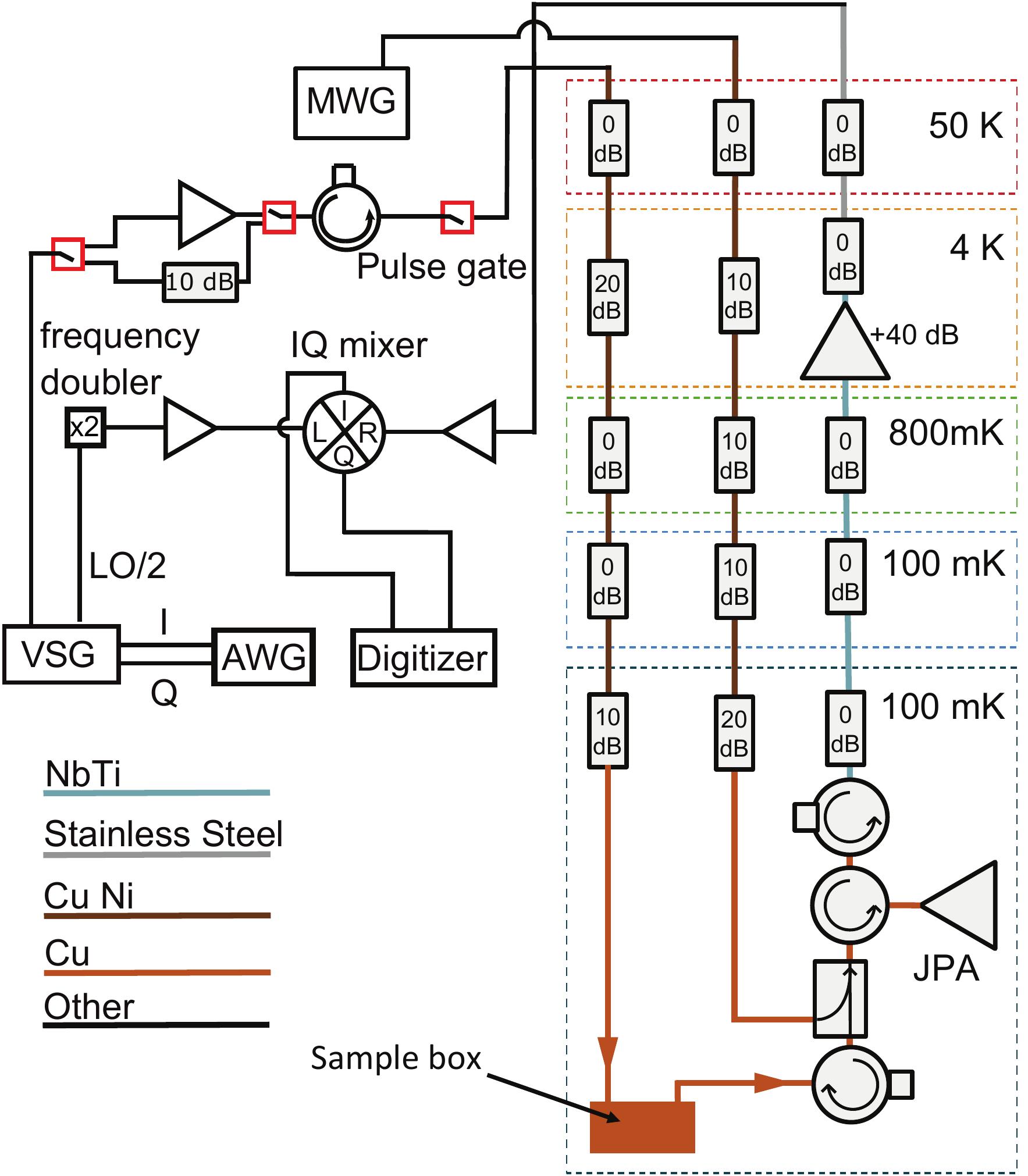}
	\caption{Schematic of the spectrometer and dilution fridge setup. Microwave pulse signals are generated using a vector signal generator (VSG) and an arbitrary waveform generator (AWG). A microwave generator (MWG) is used to drive the Josephson parametric amplifier (JPA). The return signals are down-converted and detected with a digitizer. A signal at half the up-conversion local oscillator (LO) frequency provided by the VSG is doubled with a frequency doubler and used for down-conversion. Coaxial cable types inside the fridge have been colour coded.\label{fig:Setup_full}}
\end{figure}

\subsection{Microwave measurement}
The measurement setup is shown in Fig.~\ref{fig:Setup_full}. The copper sample box is mounted at the base plate of a dilution refrigerator at 100 mK. The output signal is routed through two circulators to a Josephson parametric amplifier (JPA), a quantum limited amplifier, which amplifies the signal in reflection. The amplifier is driven via a directional coupler by a dedicated microwave source. Calibration of the JPA is given in the next section. The signal was further amplifed at 4~K by a high electron mobility transistor (HEMT) and again at room temperature before being mixed down in frequency by an IQ mixer and detected using a digitiser.

Pulsed microwave signals are generated by an arbitrary waveform generator (AWG), connected to a vector signal generator (VSG). These pulses are routed through either a high gain path (solid state amplifier +35~dB 3W max output) or an attenuating path (10~dB attenuation) both signal routes are actively gated using fast microwave switches. Signals are routed through the fridge before being mixed down at room temperature. 

\begin{figure}
\begin{subfigure}
    \centering
    \includegraphics[width=12cm]{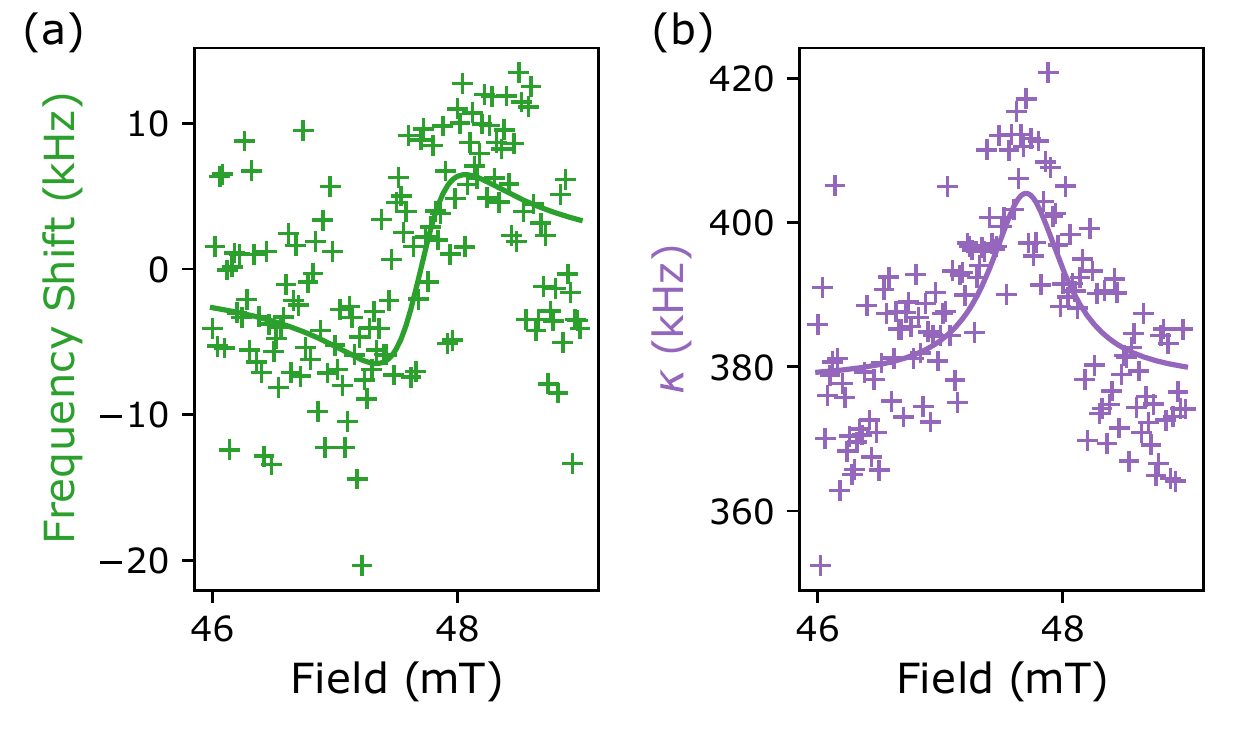}
    \label{fig:fanofit}
\end{subfigure}
%\newline
\begin{subfigure}
    \centering
    \includegraphics[width=12cm]{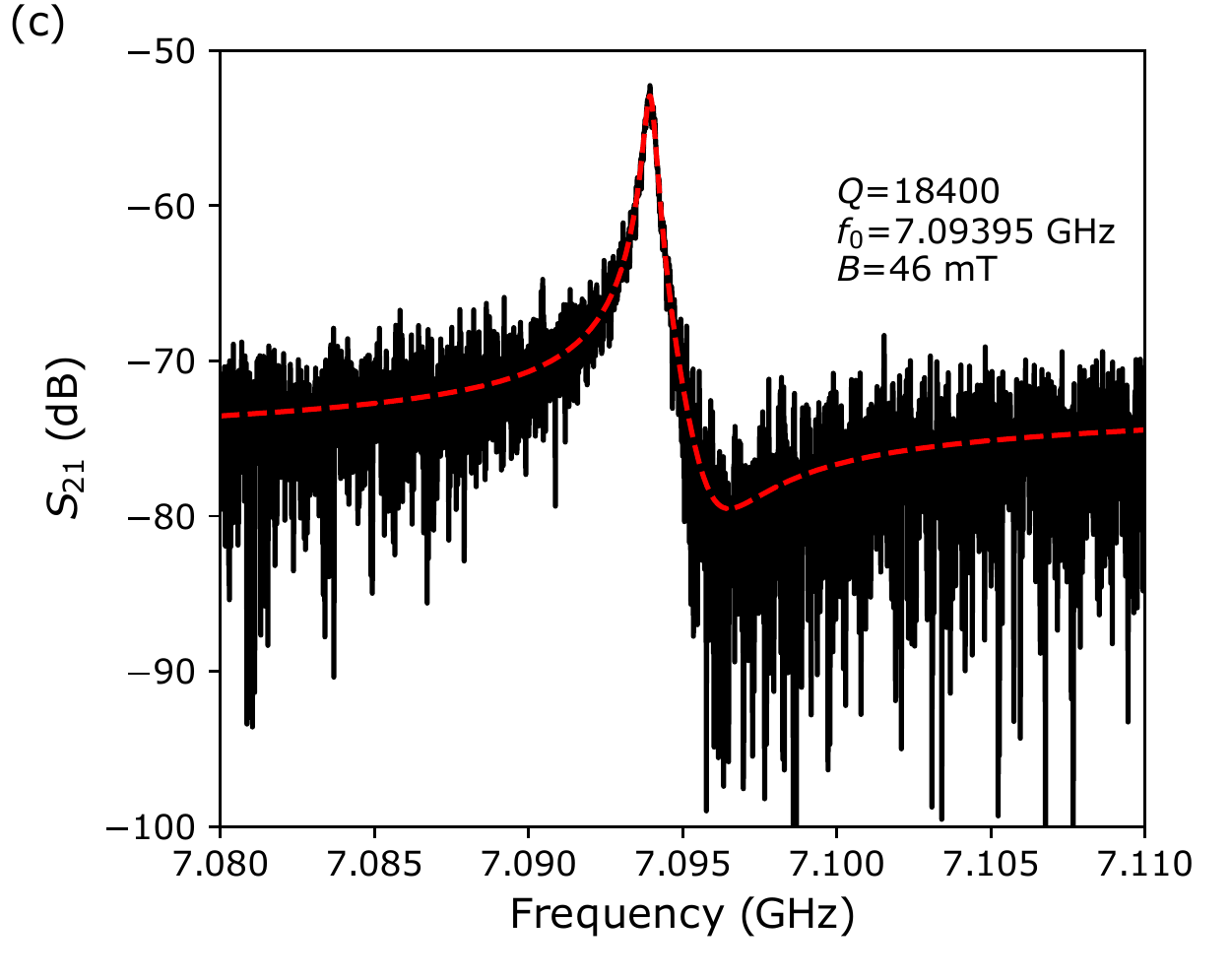}
    \caption{Effect of spins on (a) resonator frequency and (b) resonator linewidth $\kappa$ as a function of magnetic field as the spin line passes through the resonator. Data are fit simultaneously to equations in Ref.~\cite{Eisuke2011electron} $g_{\rm ens}\sim$120~kHz and $\gamma\sim2.4$~MHz. (c) Measurement of transmission magnitude through the copper cavity using a VNA at 46mT (off resonance with spin line). The superconducting resonator gives a characteristic Fano resonance response which we fit to extract a centre frequency for this resonator of 7.09395~GHz and a Q factor of 18400. }
    \label{fig:VNA}
\end{subfigure}
\end{figure}

Continuous wave measurements are performed using a vector network analyser (VNA). We measured $S_{21}$ transmission  between the antennae which is modulated by the microresonator. To characterize the resonator quality factor we fit the modulation (in linear magnitude) with a Breit-Wigner-Fano function:
\begin{equation}
    S_{\rm 21 Lin}(f) = K\frac{q\kappa/2 + f - f_0}{\kappa^2/4 + (f - f_0)^2} + mx + c
\end{equation}
where $K$ determines the size of the modulation, $q$ is an asymmetry parameter, $f_0$ is the central frequency of the resonator, $\kappa$ is the FWHM of the cavity and the $mx+c$ term is an approximation to the background transmission. This fits our resonance notch well and an example when the resonator is off resonance with the spins is shown in Fig.~\ref{fig:VNA}(c) 

\subsection{System Calibration}

\begin{figure}
\begin{subfigure}
    \centering
    \includegraphics[width=10cm]{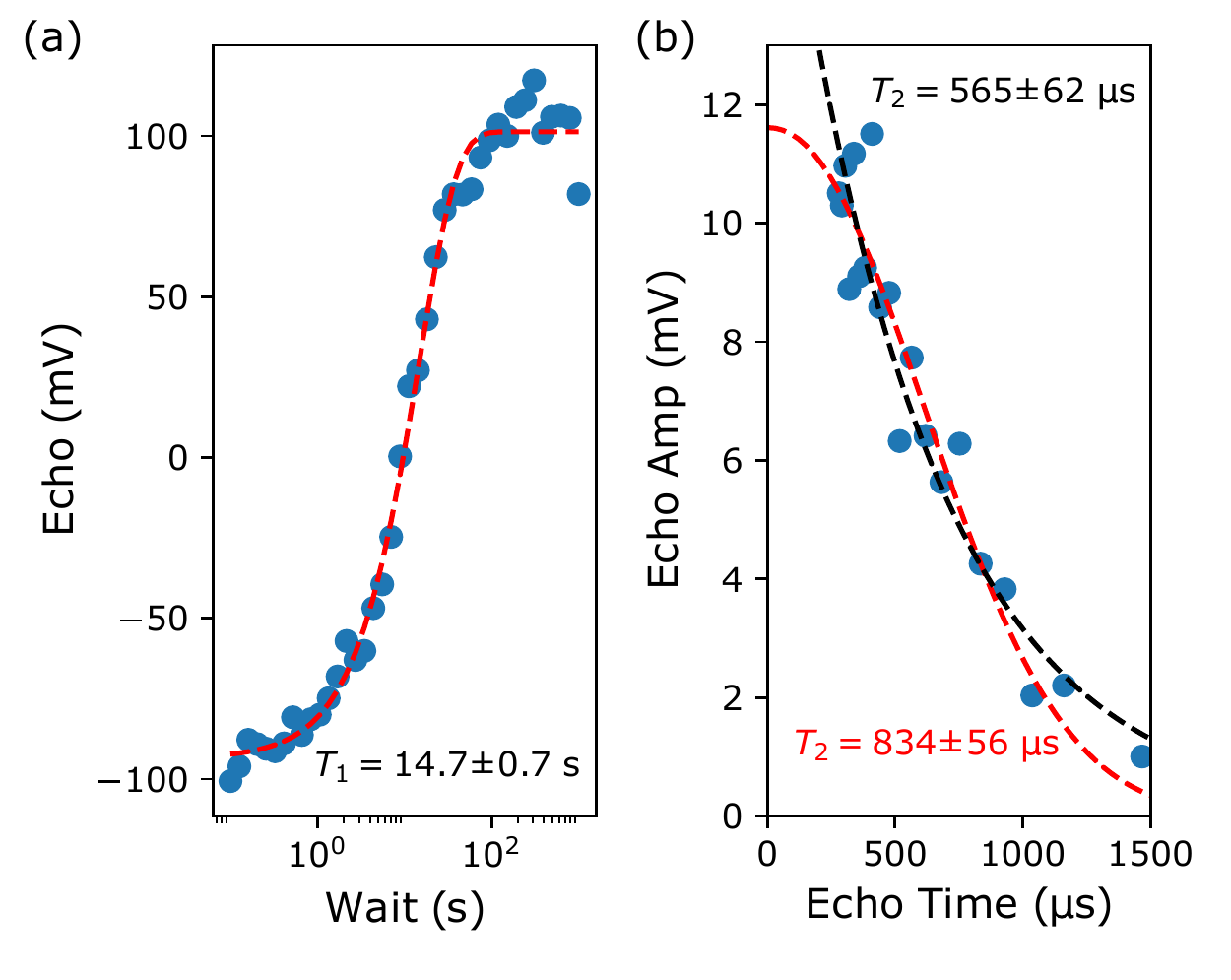}

    \label{fig:T1T2}
\end{subfigure}
%\newline
\begin{subfigure}
    \centering
    \includegraphics[width=10cm]{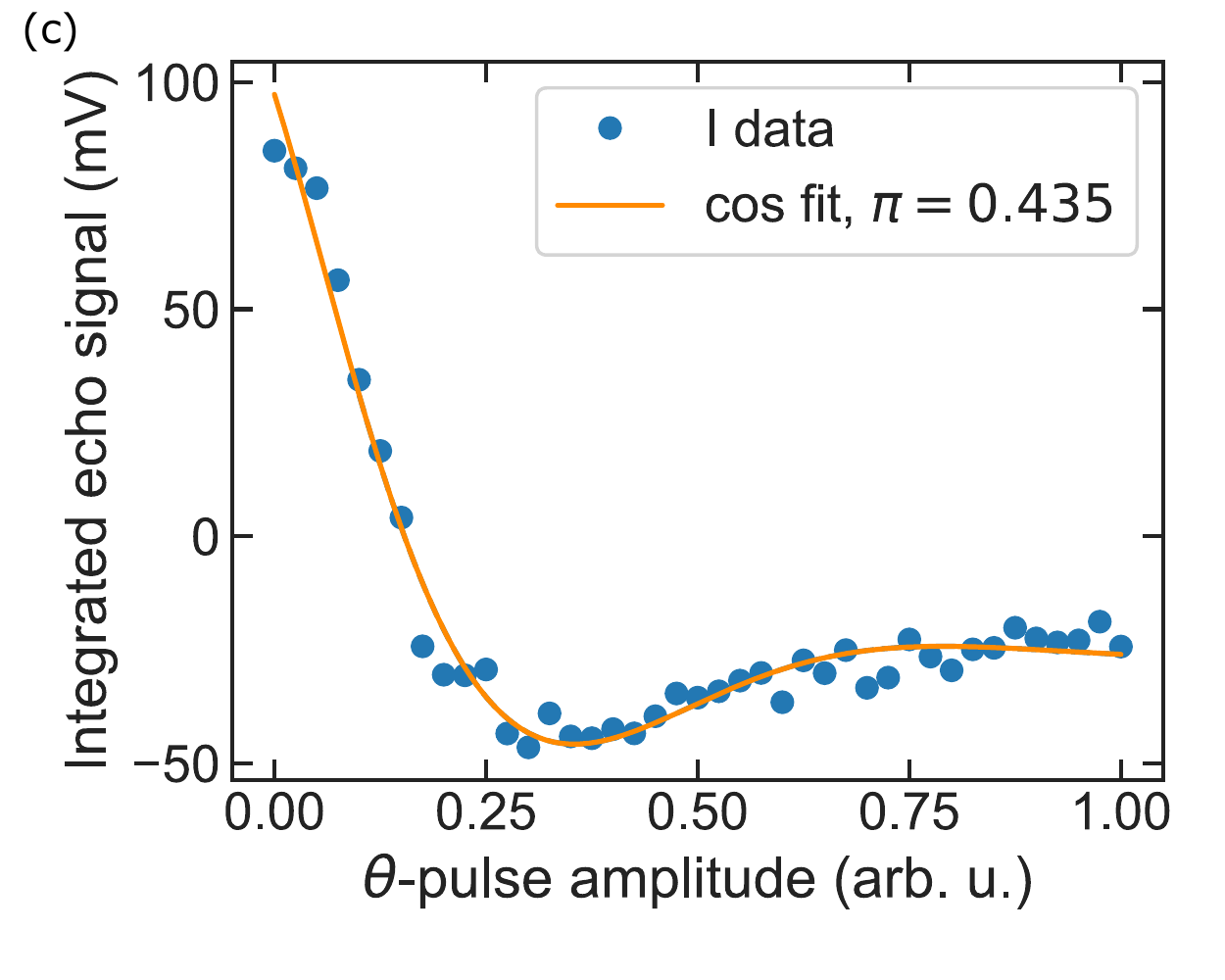}
    \caption{(a) $T_1$ by inversion recovery. A WURST pulse is used to invert the ensemble before a wait (indicated on the x axis) and a detection sequence at constant high power and $\tau$. (b) $T_2$ measurement of the sample at the same weak excitation strength as used in Fig.~2 main text. The pulse sequence (inset) is a weak excitation followed by two 100~$\upmu$s WURST pulses. Increasing $\tau$ changes the echo time allowing $T_2$ to be measured. Fitting stretched or single exponentials gives different $T_2$ giving a combined uncertainty of $T_2 = 0.7(2)$~ms (c) Rabi oscillations using a Gaussian $\theta$-pulse of duration 8~$\upmu$s and FWHM 4~$\upmu$s and a two-WURST silenced echo detection sequence. The oscillations are heavily damped due to a large inhomogeneity in Rabi frequency across the ensemble.}
    \label{fig:Rabi}
\end{subfigure}
\end{figure}

We measure $T_1$, $T_2$ and the Rabi frequency of the hybrid system and show these measurements in Fig.~\ref{fig:T1T2}. $T_1$ is measured by inversion recovery where the WURST pulse inverts the spin ensemble and we measure the time taken for the inverted ensemble to return to the ground state and find $T_1\sim$14~s. $T_2$ is measured using two WURST pulses to refocus a weak excitation (same strength as in Fig.~2). Varying $\tau$ allows us to measure the effect of dephasing on the echo amplitude. We fit a single quadrature of the echo decay returning a coherence time of 0.7(2)~ms where most of the uncertainty arises due to different values returned fitting a stretched or single exponential to the data.

To measure the Rabi frequency, $\Omega$, we appply a  $\theta$ rotation pulse followed 10~ms later by a detection sequence with VSG output power of -20~dBm sent through the high gain path. The $\theta$-pulse is chosen to match the duration and shape of the excitations used in the memory sequence in Fig.~3. The heavily damped envelope to sinusoidal Rabi oscillations (shown in Fig.~\ref{fig:Rabi}) is due to a large inhomogeneity in single spin coupling, $g_0$, across the ensemble and introduces substantial ($\sim 50\%$) uncertainty to this measurement. We fit a decaying cosine function to the echo amplitude, allowing us to extract an approximate $\pi$ pulse amplitude of 0.435 (where 1 is the maximum pulse amplitude at the chosen VSG output power) giving $\Omega = 125$~kHz. 

Using $T_1$ and $\Omega$ we calibrate the photon number in the resonator. The Purcell effect limits $1/T_1 = 4g_0^2/\kappa$, allowing a measure of average spin-resonator coupling $g_0$. The different $\kappa$ dependence from Ref.~\cite{bienfait2016controlling} is due to the definition of $\kappa$ being the HWHM in this work and FWHM in Ref.~\cite{bienfait2016controlling}. At the centre of the line $\kappa\sim400$~kHz, $T_1 = 14.7$~s giving $g_0\sim~80$~Hz ($g_0$ varies across the ensemble and this number is indicative). The Rabi frequency can be written as $\Omega = 2g_0\sqrt{n}$ which allows us to calibrate the photon number $n\sim5.7\times10^5$ for the $\pi$ pulse above. We can rescale this to the photon number to the memory sequence in Fig.~3  (main text) based on an amplitude of 0.2 at a power $\sim$20~dB lower than the Rabi measurement and find $\langle n\rangle \sim$ 1200 photons for the memory protocol. Due to the uncertainty in Rabi measurements these photon numbers should be treated as indicative powers accurate to approximately $50\%$.

We note that in Fig.~2(d) main text, that repeated WURST pulses before a pulse sequence can cause a reduction in echo amplitude which implies that WURST pulses cause some saturation of the ensemble. However, comparing the $T_2$ decay shown here and the $T_M$ (Fig.~2(e) main text) acquired when using repeated WURST pulses, we see that repeated WURST pulses actually increase the final echo amplitude due to dynamical decoupling. These two observations appear initially at odds. However, we understand this discrepancy using a toy model where we divide spins into three zones, (i) weakly coupled spins, far from the resonator and unaffected by WURST pulses, (ii) spins a moderate distance from the resonator, in turn moderately coupled and undergo imperfect inversion under WURST pulses (iii) spins close and strongly coupled to the resonator which are inverted by WURST pulses with high fidelity. Repeated WURST pulses will saturate zone (ii) which would contribute some of the echo strength. When operating this memory with a long string of WURST pulses, zone (ii) would always remain saturated, and therefore not contribute to the protocol. This means that in future devices, sufficient cooperativity for high fidelity memory transfer must be achieved with only spins in zone (iii).

\begin{figure}
\begin{subfigure}
    \centering
    \includegraphics[width=12cm]{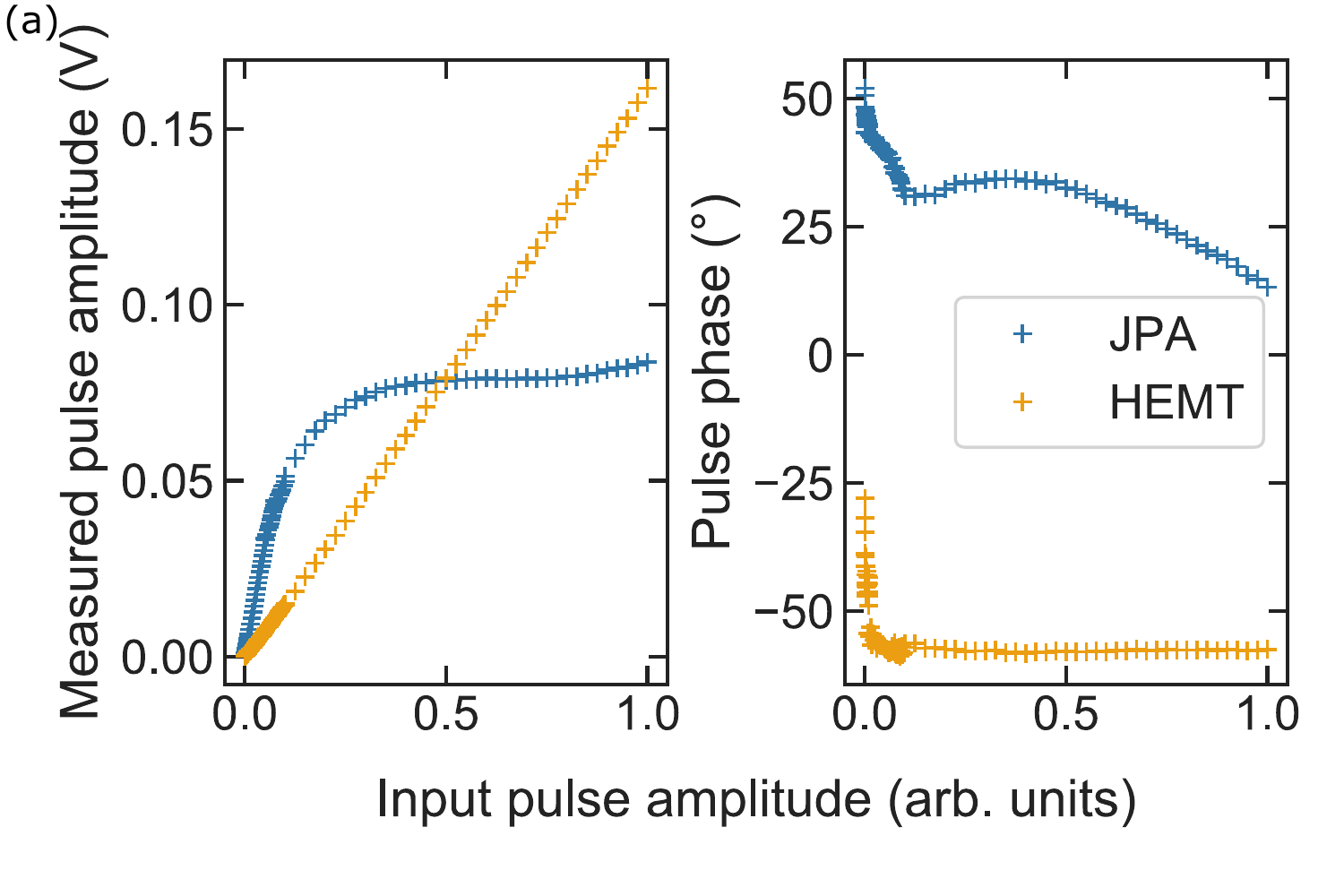}

    \label{fig:JPA_cal}
\end{subfigure}
%\newline
\begin{subfigure}
    \centering
    \includegraphics[width=12cm]{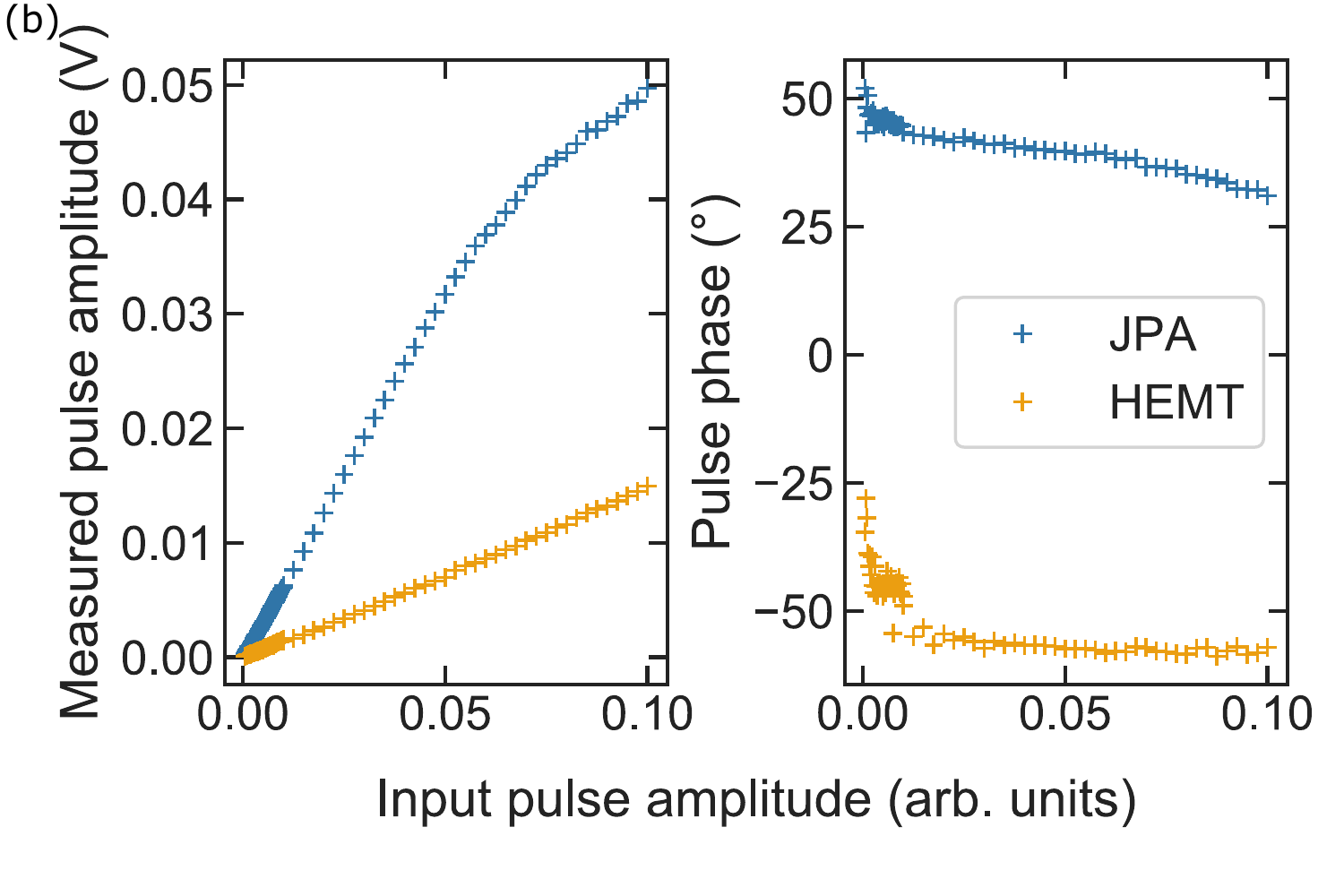}
    \caption{(a) Measured pulse amplitude as a function of input pulse amplitude, measured using the JPA (blue) and the HEMT without JPA (orange). (b) Low power regime of the same experiment.}
    \label{fig:JPA_cal_lowpower}
\end{subfigure}
\end{figure}

The use of a JPA in measurements was essential to perform experiments at low photon numbers. Such amplifiers are prone to saturation and nonlinearity with large input signals. We calibrate the JPA to determine the onset and extent of the nonlinear regime, and any additional distortions that may be present. For comparison, the same measurements were repeated with the JPA turned off, using only a HEMT at 4~K for amplification. We confirm that for Fig.~3  (main text)the echo signals are well into the linear regime but the input excitations are in the nonlinear regime of the JPA and are prone to distortion. The HEMT remains linear and un-distorted at all powers, as expected. JPA calibrations are shown in Fig.~\ref{fig:JPA_cal_lowpower}. The inhomogeneous linewidth of the ensemble is $\sim$2.5~MHz (see Ref.~\cite{o2020spin} for further details). 

To further increase signal we chose an operating temperature to maximise the population of the ground state of our chosen bismuth transition ($\sim$10~\%). Pulsed experiments are run at long a shot-repetition rate of 160~s to minimise the saturation of spins. These considerations are to improve the cooperativity between the spin ensemble and resonator which contributes to the read/write efficiency of the memory and thus improve signal - approximately 17~\% one way efficiency as shown in the main text. This means that when measuring $\sim$200 photon input pulses, the output pulse is actually $0.17^2\times200 \sim 6$ photon output pulses - without considering any $T_2$ decay. Despite our optimized antenna configuration, we may fail to capture all of this output pulse, and also find that the JPA is working slightly sub-optimally (given the elevated temperature). The combination of these effects means that to observe a low-noise echo at low powers we typically required hundreds of averages. Recent work using single microwave photon detectors~\cite{albertinale2021detecting} is a promising route to improve sensitivity. In tandem it is necessary to consider routes to improve memory efficiency in order to achieve useful memories. Promising approaches to this include the use of nuclear-spin 1/2 species (allowing the ensemble to be 100~\% thermally polarised at mK), optimal resonator design, increasing filling factor whilst maintaining large single-spin coupling necessary for WURST inversion pulses and the use of $^{28}$Si substrates which would improve coherence times and also narrow inhomogeneous spin linewidths. 

\subsection{Cooperativity}
Cooperativity in a hybrid micro-resonator spin system can be measured by the dispersive shift to the resonator frequency ($f$), and the increased half-width half max of the resonator frequency response ($\kappa$) \cite{Eisuke2011electron}:
\begin{equation}
    f= f_0 - \frac{g_{\rm ens}^2\Delta}{\Delta^2 + \gamma^2/4}
    \label{eq:freq}
\end{equation}
\begin{equation}
    \kappa = \kappa_0 + \frac{g_{\rm ens}^2\gamma/2}{\Delta^2 + \gamma^/4}
    \label{eq:kappa}
\end{equation}
where $\kappa_0$ and $f_0$ are respectively the half width and frequency of the resonator in the absence of spins, and the detuning $\Delta = (B_0 - B_R)\times \partial f/\partial B_0$, where $B_R$ is the magnetic field at which spins and resonator are resonant. Fitting equations \ref{eq:freq} and \ref{eq:kappa} to our data gives $C\sim0.067$ as shown in Fig.~\ref{fig:VNA}.

We also extract the cooperativity from $T_2$ measurements with CPMG sequences in Fig.~2  (main text) by modelling the echo intensity as a function of echo number and considering the reduction of energy in the echo field with repeated emission. In A[BB]$_n$A sequences the echo is emitted after the final A pulse. As (i) WURST pulses efficiently refocus spins and (ii) only one echo forms, we attribute any change in echo intensity to dephasing of the spins. We fit the echo intensity  $A_{\rm sil}(t)$ to extract the dephasing time $T_2$ using
\begin{equation}
    A_{\rm sil}(t) = \sqrt{A_0^2 \exp(-2t/T_2) + K^2},
    \label{eq:T2_sil}
\end{equation}
where $A_0$ is the echo amplitude at time $t=0$, $T_2$ is the dephasing time and $K$ is the background giving a dephasing time of $T_2 = 2.0\pm 0.2$~ms for a refocusing rate of 7.1~kHz.
We model the echo amplitude in A[AA]$_n$A sequences by assuming that every time an echo forms, a constant fraction of the energy stored in the echo field is lost and that in between echo emission the echo field dephases as in  A[BB]$_n$A sequences. We model the echo amplitude by
\begin{equation}
    A(t) = \sqrt{A_0^2 \exp(-2t/T_2)\times(1-\eta_{\rm em})^{N} + K^2}
    \label{eq:T2_unsil}
\end{equation}
where $\eta_{\rm em}$ is the efficiency with which energy is emitted from the echo field to the cavity, $N$ is the number of echoes which have occurred before time $t$. This is fit with one free parameter, $\eta_{\rm em}$, and is shown shown in Fig.~2.

Using Equation (17) from Ref.~\cite{afzelius2013proposal} we relate the one-way efficiency to the cooperativity by 
\begin{equation}
    \eta_{\rm em} = \frac{4C}{(1+C)^2}
\end{equation}
giving $C=0.047$ in close agreement to the value extracted in the more typical method by measuring effects on the resonator. 

\subsection{WURST Pulses}

WURST pulses are a type of adiabatic fast passage which have been used to control spin ensembles~\cite{o2013wurst}. They have advantages in the context of semi-classical control as, in the adiabatic limit, they impart faithful $\pi$ pulses across a wide bandwidth of spins irrespective of the coupling between the spin and the drive field which may be inhomogeneous across the ensemble. By chirping the frequency of the control pulse, the effective magnetic field observed from the rotating frame of the un-driven spin flips from pointing along one pole of the Bloch sphere to pointing along the other pole following a path along the surface of the Bloch sphere. In this reference frame the spin precesses around this effective magnetic field, and if that precession rate is sufficiently high relative to the rate of the field reversal (i.e. the pulse is sufficiently strong relative to the chirp rate) then the spin will adiabatically track this magnetic field and undergo a $\pi$ inversion pulse. This is described well in section 2.2 of Ref.~\cite{o2013wurst}. In this work we use WURST pulses of order 20. 

\begin{figure}
    \centering
    \includegraphics[width=0.5\textwidth]{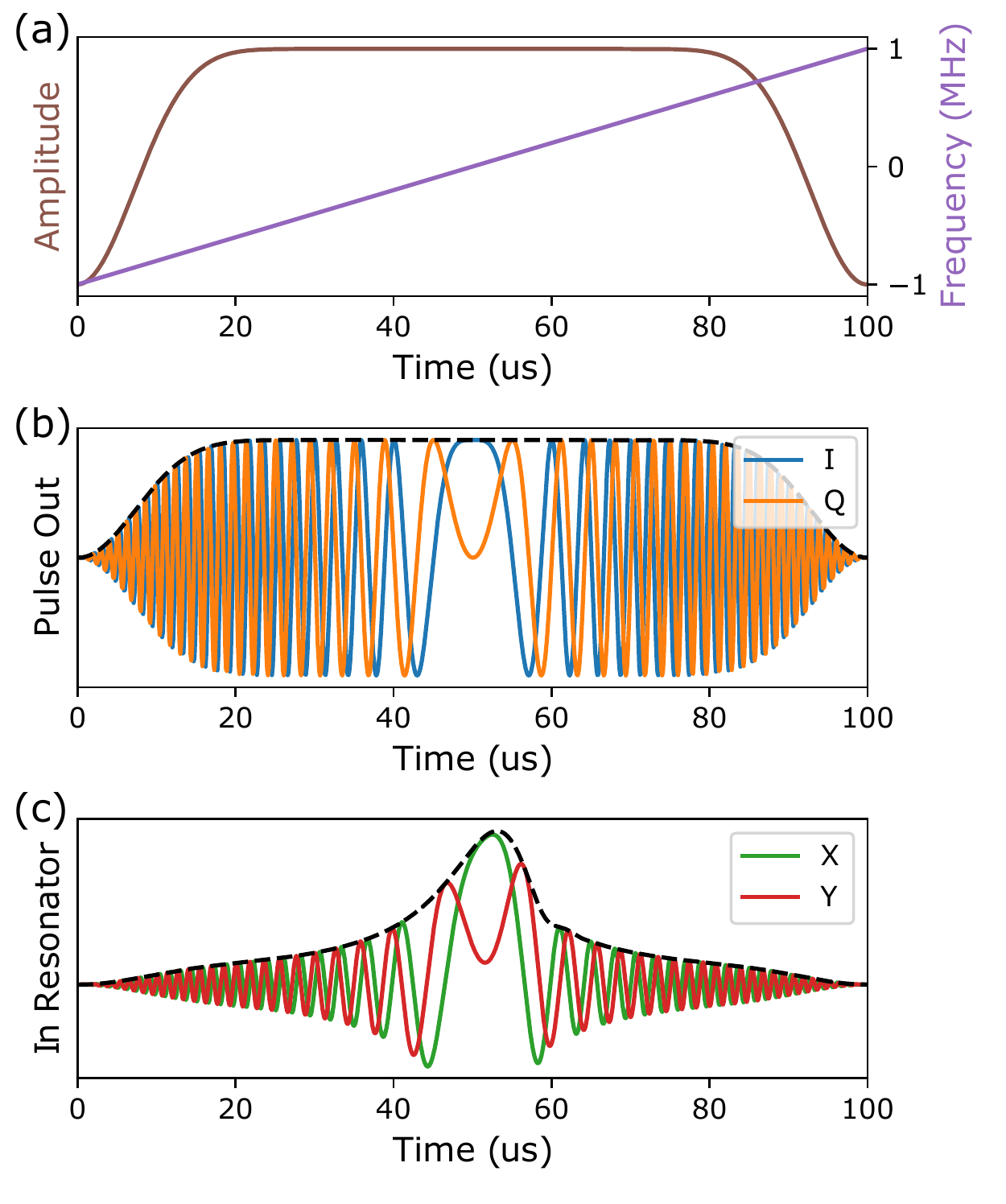}
    \caption{WURST-20 pulses with (a) the pulse amplitude and frequency shift for a WURST pulse with bandwidth 2~MHz and duration 100~$\upmu$s. (b) The I and Q quadratures of the same WURST pulse. (c) The X and Y field quadratures in the cavity once the WURST pulse is filtered by a $\kappa=$200~kHz cavity. }
    \label{fig:WURST}
\end{figure}

\subsubsection{Input Output Theory}
Wideband uniform rate smooth truncation (WURST) pulses can be characterized by frequency and amplitude modulation (FM and AM respectively)
\begin{align}
    f_{\rm W}(t) = -\frac{\Gamma_{\rm W}}{2} + \frac{\Gamma_{\rm W}}{T_{\rm W}}t,\\
    A_{\rm W}(t) = 1 - \abs{\sin(\pi\frac{t-T_{\rm W}/2}{T_{\rm W}})}^N, \nonumber
\end{align}
where $\Gamma_{\rm W}$ is the bandwidth of the WURST pulse, $T_{\rm W}$ is the WURST pulse duration and $N$ is the WURST pulse index, in this work we use $N=20$ pulses. 

The FM results in a time varying phase of the output pulse
\begin{align}
    \phi(t) = \int_0^t f_{\rm W}(t') dt' + \phi_0,
    \label{eq:wurst_phi}
\end{align}
where $\phi_0$ is the phase of the WURST pulse. The I and Q pulse quadratures applied by the AWG are
\begin{align}
    I(t) = A_{\rm W}(t) \sin\phi(t),\nonumber\\
    Q(t) = A_{\rm W}(t) \cos\phi(t),
\end{align}
The WURST pulse is modulated by the cavity described by input/output theory \cite{gardiner1985input, ranjan2020pulsed} where 
\begin{align}
    \dot{X}(t) = \sqrt{\kappa_C}I(t) - \frac{\kappa}{2} X(t),\nonumber\\
    \dot{Y}(t) = \sqrt{\kappa_C}Q(t) - \frac{\kappa}{2} Y(t),
\end{align}
where $\kappa_C$ is the coupling linewidth of the cavity and $\kappa$ is the loaded linewidth of the cavity. Increasing the bandwidth of the WURST pulse above the bandwidth of the cavity shortens the effective WURST pulse duration. We show an example WURST pulse in FM/AM (Fig.~\ref{fig:WURST}a), consequent IQ quadratures (Fig.~\ref{fig:WURST}b) and after modulation by a cavity (Fig.~\ref{fig:WURST}c)
 
\subsubsection{Limits on WURST Pulses}
\begin{figure}
    \centering
    \includegraphics[width=0.5\textwidth]{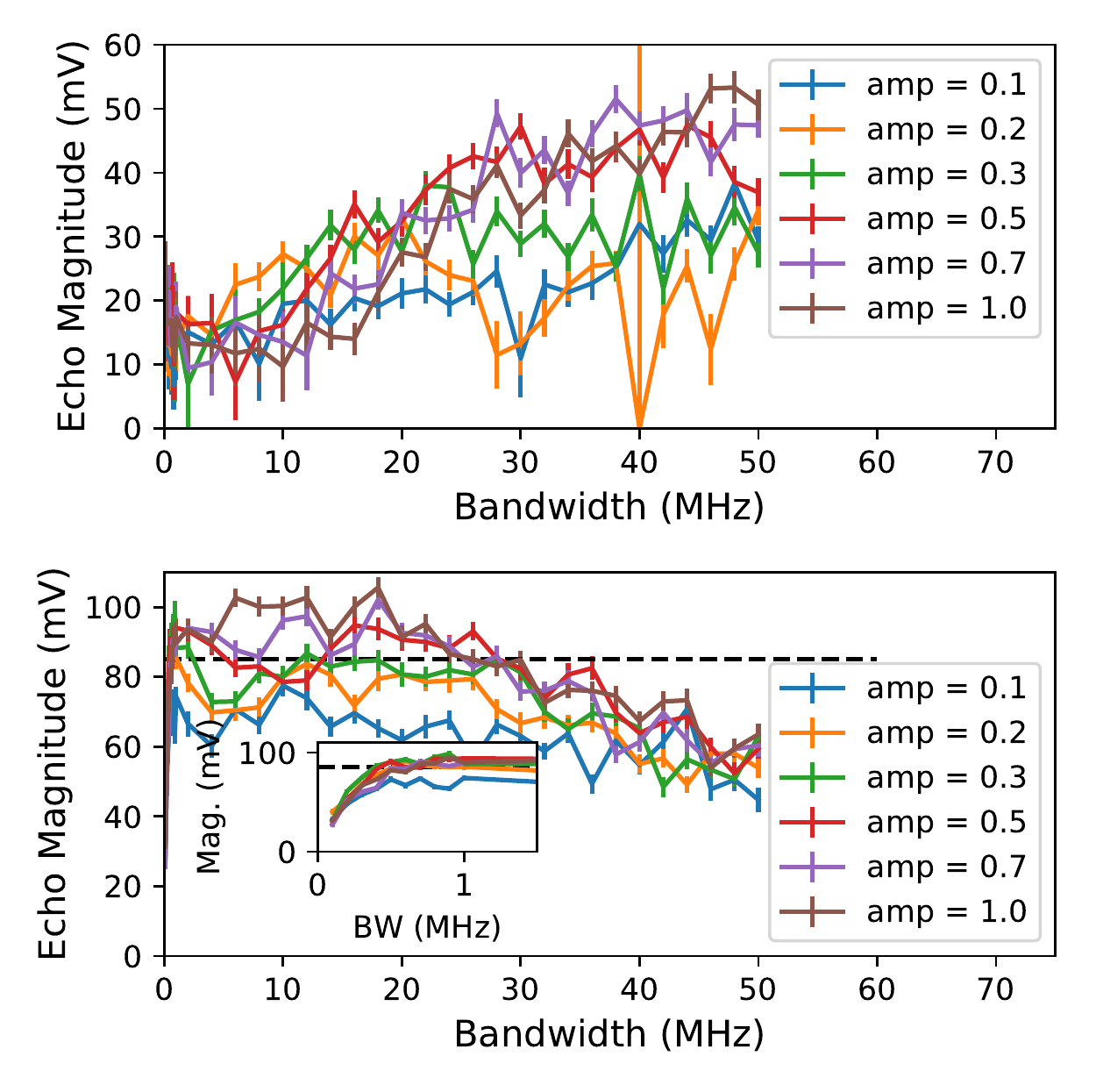}
    \caption{Top (bottom) panel – amplitude of the first (second), nominally silenced (unsilenced), echo as a function of the pulse bandwidth for 200us WURST pulses. Inset to the bottom panel is the unsilenced data for low bandwidths, showing that there is a minimum bandwidth, similar to the cavity width, required to maximise the echo amplitude by refocusing all the spins in the cavity. The adiabaticity limit is determined by thresholding the unsilenced echo intensity relative to 85~mV shown as a black dashed line. Non-monotocity of echo intensity and uncertainty in echo amplitude give uncertainty in the thresolded value shown in Fig.~4.}
    \label{fig:silencing}
\end{figure}

In Fig.~4, a region of suitable WURST pulses --- WURST pulses which inverts a large fraction of the spin ensemble --- is mapped out based on both theory and experiment described in this section. Using pulse sequences such as that in Fig.~1(a) (excitation followed by two identical WURST pulses) we have measured the refocusing efficiency. We measure the amplitude of the first (nominally silenced) and second echo and present them as a function of the WURST pulse bandwidth and amplitude in Fig.~\ref{fig:silencing}. A `good' WURST pulse results in no echo after the first WURST, and a loud echo after the second. 

By thresholding the lines we determine regions where WURST pulses adiabatically refocus a large fraction of spins. The constraint that the second echo must have a suitable magnitude (85~mV) gives the experimental boundary to suitable WURST pulses shown in Fig.~4  (main text) based on adiabaticity. 

This is supported by a theoretical treatment~\cite{doll2013adiabatic}.
\begin{equation}
    Q_{\rm min} = 2\pi\nu^2/R \gg 1,
\end{equation}
where $ Q_{\rm min}$ is an adiabaticity factor, $R$ is the chirp rate and $\nu$ is the nutation frequency (how quickly the spin precesses around the effective field). $\nu$ is minimal (and equal to the Rabi frequency) when the WURST frequency is the same as the spin frequency.
We choose a minimum $g_0$ for spins to be refocused and solve this equation for $Q_{\rm min} = 1$ finding a lower limit on pulse strength. 

In the homodyne scheme used in this work there is a limit placed by the finite frequency resolution we can imprint onto WURST pulses. The effective duration of the WURST pulse, $T_{\rm W, eff} = \kappa T_{\rm W}/\Gamma_{\rm W}$, is the time the WURST pulse of duration $T_{\rm W}$ and total bandwidth $\Gamma_{\rm W}$ takes to chirp across the cavity frequency $\kappa$. The inverse of this time gives an indication of the minimum frequency that can be resolved by modulating the pulse. This frequency must be less than the cavity bandwidth placing a limit on the maximum $\Gamma_{\rm W}$ which can be used,
\begin{equation}
    \Gamma_{\rm W} \ll \kappa^2 T_{\rm W} = 37~\mathrm{MHz}.
\end{equation}

The maximum spectrometer power limits the power of WURSTs we can apply and is shown as the top boundary in Fig.~4. This line cannot be increased to arbitrarily high powers using higher power amplifiers as the superconducting resonators will limit the maximum power. There is also a minimum WURST bandwidth imposed by the cavity bandwidth. 

\subsubsection{WURST pulse distinctiveness}

\begin{figure}
\begin{subfigure}
    \centering
    \includegraphics[width=10cm]{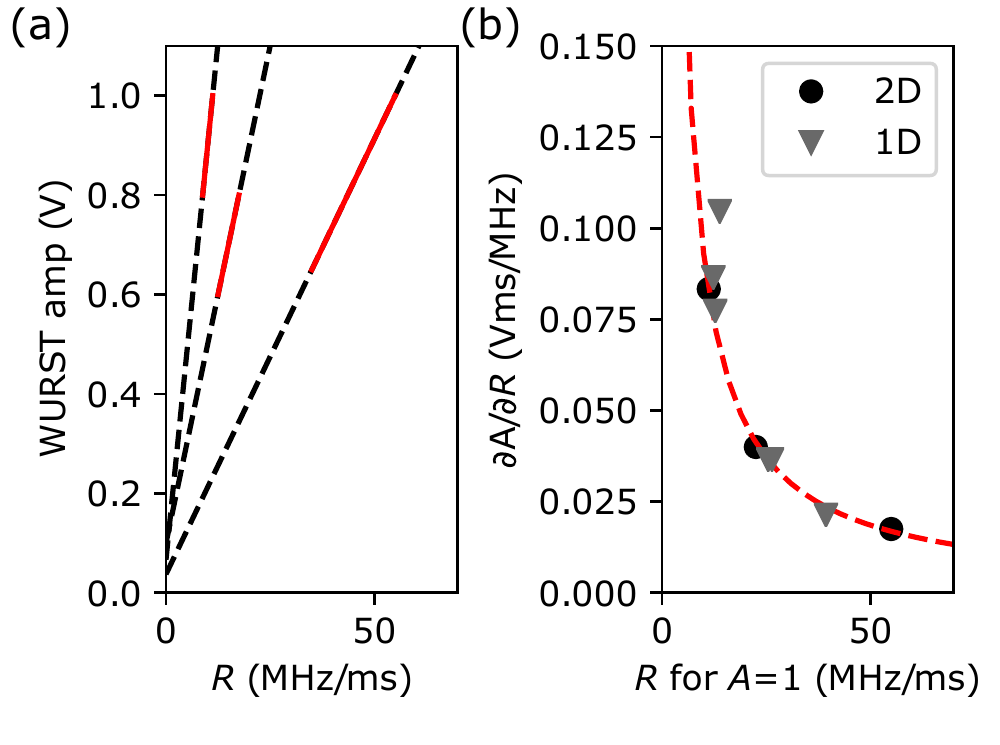}
    \label{fig:gradients}
\end{subfigure}
%\newline
\begin{subfigure}
    \centering
    \includegraphics[width=10cm]{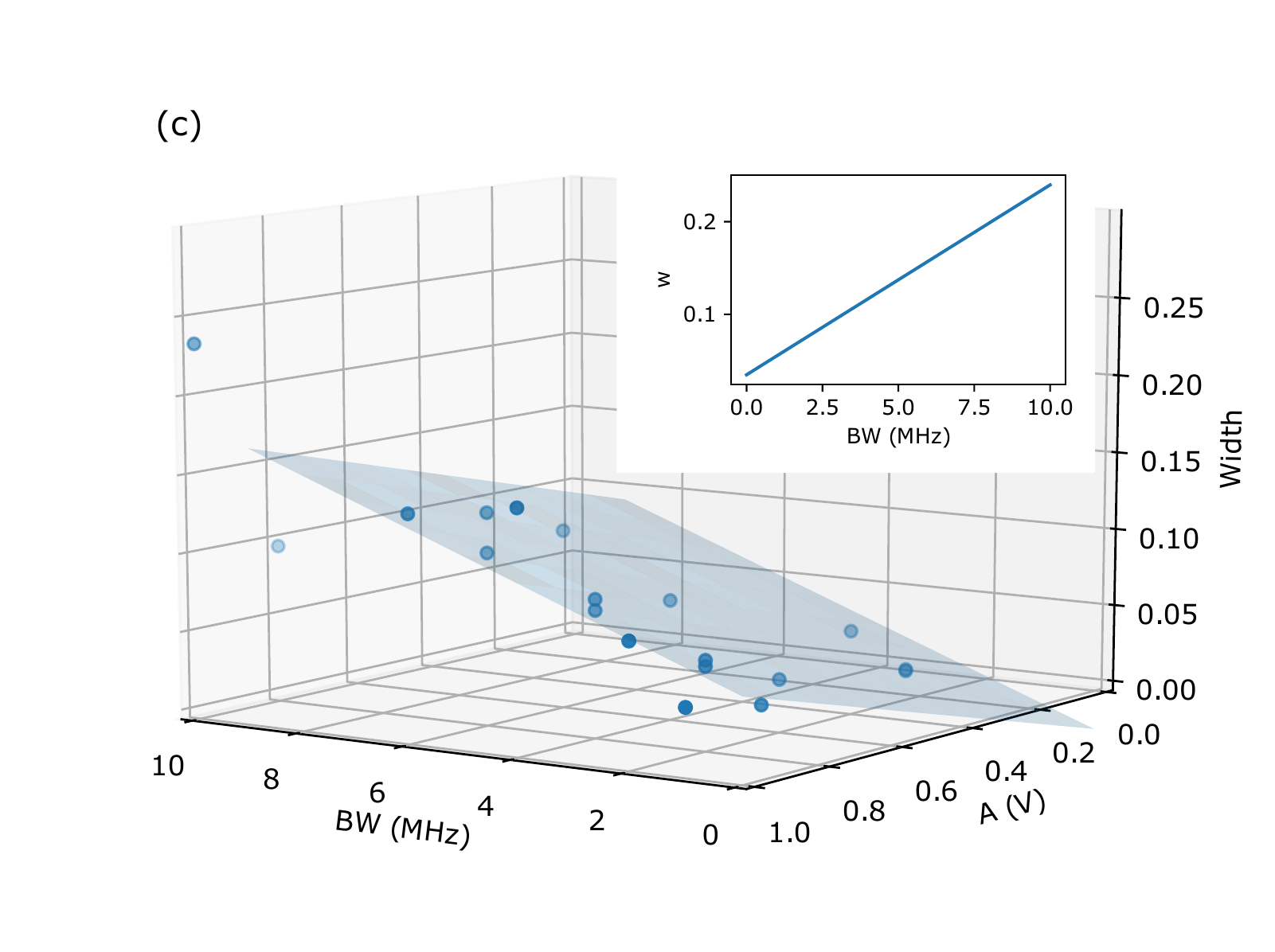}
    \caption{(a) The line of strong echoes from 2D maps of AB-echo sequences. We fit a line to these data and extract their gradient. (b) The gradient of the lines fit in (a) as a function of their extrapolated bandwidth at amplitude $A$=1 (black circles). The gradient determined when B-pulse has been swept in 1D in both amplitude and bandwidth are inferred and plot on the same axes (grey triangles) showing good agreement. We fit these data with $\partial A/\partial R = C/R$ with $C$ a fit parameter and return the red dashed line and $C=0.93$. (c) Width of the line of equivalent WURST pulses as a function of WURST pulse parameters. We fit a plane to this data to allow interpolation of the width as a function of bandwidth and amplitude.}
    \label{fig:widths}
\end{subfigure}
\end{figure}

We partition the space of suitable WURSTs into `distinct' WURST pulses to estimate the memory capacity. We partition this region based upon two-WURST echo sequences where the first and second WURST pulse have different properties. Inset to Fig.~4  (main text) is the echo amplitude as the amplitude and chirp rate of the B pulse is varied. This sequence results in strong echoes along a line with positive gradient. In the main text we explain that this line of equivalent pulses is due to the acquisition of a dynamic phase. Together with other two dimensional and one dimensional maps varying a B-pulse parameter we interpolate the gradient and width of the line of equivalent pulses across the space of WURST pulses giving good inversion and use this interpolation to count WURST pulses. This interpolation is shown in Fig.~\ref{fig:widths} with the results in Fig.~4.  

\subsubsection{Theoretical description of WURST pulses}
The cavity is driven by a chirped field which results in a temporal intensity profile of the field inside the cavity, and the coupled equations for the spins and the cavity field must in general be solved numerically. Here we shall make some observations for a simplified model, that will qualitatively explain most features seen in the experiments. The spins have a central frequency $\omega_c$ and each spin is characterized by its detuning $\delta=\omega-\omega_c$.
We characterize a WURST pulse by its chirp rate $R$ and pulse centre $t_0$, yielding its frequency $\omega(t) = R (t-t_0)$ and phase $\phi(t) = \frac{1}{2}R (t-t_0)^2$ in the frame rotating at $\omega_c$. In that frame the Hamiltonian for each given spin is $\delta \sigma_{z}/2$, and the time dependent complex Rabi frequency is $\Omega(t)= \Omega_0 \exp(-i\phi(t))$.

The spin transitions are resonant when $R(t-t_0)=\delta$, i.e., at the time $t_\delta\equiv t_0 + \delta/R$. In the interaction picture with respect to the spin excitation energy $\delta \sigma_{z}/2 $, the Hamiltonian of a single spin reads,
 \begin{equation} \label{eq:HSI}
H_I(t)= \frac{\Omega_0}{2} ( e^{-i\phi(t)} e^{i\delta t} \sigma^+ + e^{i\phi(t)} e^{-i\delta t} \sigma^- )
\end{equation}
where $\phi(t) - \delta t = \frac{1}{2} R(t-t_\delta)^2 + \delta t_0 - \delta^2/2R$. 

Assuming a near perfect adiabatic chirp, the unitary evolution operator of the full chirp process is given in the $\sigma_z$ eigenstate basis:
\begin{equation} \label{eq:U}
U=\left(
    \begin{array}{cc}
      0 & -i e^{-i\theta_\delta} \\
      -ie^{i\theta_\delta} & 0 \\
    \end{array}
  \right)
\end{equation}
where $\theta_\delta = \phi_W -\delta t_0$. $\phi_W$ contains a term $\delta^2/2R$ and a term that does not depend on $\delta$ but will in general depend on the chirp rate $R$ and on the coupling strength $\Omega_0$ (see below).

For initial spin states with a (small)  excitation amplitude of $\epsilon$ at $t=0$ , the $\delta$-dependence of the mean value of the $\sigma^+$ operator is $\propto \epsilon e^{2i\delta t_0}$ (disregarding the contribution from $\phi_W$), and in the Schr\"odinger picture, multiplying by $e^{-i\delta t}$ we obtain $\sigma^+ \propto  \epsilon e^{i\delta (2 t_0 - t)}$. If it was not for the detuning dependence of $\phi_W$, these terms would rephase at the time $t=2t_0$ as in the conventional $\pi-$ pulse echo. Now, instead, the spins may have very different excitation phases and the integral over $\delta$ vanishes and the echo is silenced.

All WURST pulses are described by the same form \eqref{eq:U}, and the action of two subsequent pulses (in the interaction picture) at times $t_0$ and $t_1$ is readily found,
\begin{equation} \label{eq:U2}
 U = U_1  U_0 =
  \left(
      \begin{array}{cc}
             -a^* b  &  0 \\
     0 &  -ab^* \\
    \end{array}
  \right)
\end{equation}

With $a=e^{-i\theta_{\delta}}$, with parameters $R_0, \Omega_0$ and $b=e^{-i\theta_\delta}$, with parameters $R_1, \Omega_1$, we get
$ab^*=e^{i(\phi_{W_0}-\phi_{W_1})+i\delta(t_1-t_0)}$ and if
the pulse parameters are identical, this simplifies to $ab^*=e^{i\delta(t_0-t_1)}$.
Thus, after two identical WURST pulses, a weak spin excitation returns with a global minus sign and a detuning dependent phase factor. In the Schr\"odinger picture, the linear detuning dependence of the phase disappears at $t=2(t_1-t_0)$ and we obtain an echo.

Assuming perfect inversion, a sequence with an even number of WURST pulses yields a final phase which is the sum of all WURST phases $\theta_\delta$ with $+(-)$ signs if they are raising (lowering) pulses, cf., \eqref{eq:HSI}.
Two pulses with different chirp rates or strengths will not cause an echo due to the nontrivial dependence of the phase difference $\phi_{W_0}-\phi_{W_1}$. The quadratic variation $\delta^2(1/2R_0 - 1/2R_1)$ due to the phase chirp may thus cause destructive interference over the distribution of $\delta$ and the dynamical phase associated with the energies of the adiabatic eigenstates during the chirped adiabatic process may cause destructive interference over a range of values for the coupling strength.

As we observe  in the experiments, variation of the chirp rate and strength permit addressing of separate storage modes. The field amplitude $\Omega_0$, and the chirp rate $R$  both appear in the dressed state energies and hence in the dynamical phases, and the two control parameters turn out not lead to exploration of a two dimensional space of storage modes . An analytical theory  for the phase textures indicated in Fig.1(b) would require solution of the driven two-level dynamics by a chirped frequency field with a time and frequency dependent Rabi frequency $\Omega_0(t)$ imposed by the finite bandwidth cavity mode. While this is not possible, we can solve the dynamics numerically and do so to elucidate two effects shown in Fig.~\ref{fig:theory_efficiency} and Fig.~\ref{F2} respectively.

\subsection{High power memory sequences}
\begin{figure}
    \centering
    \includegraphics[width=\linewidth]{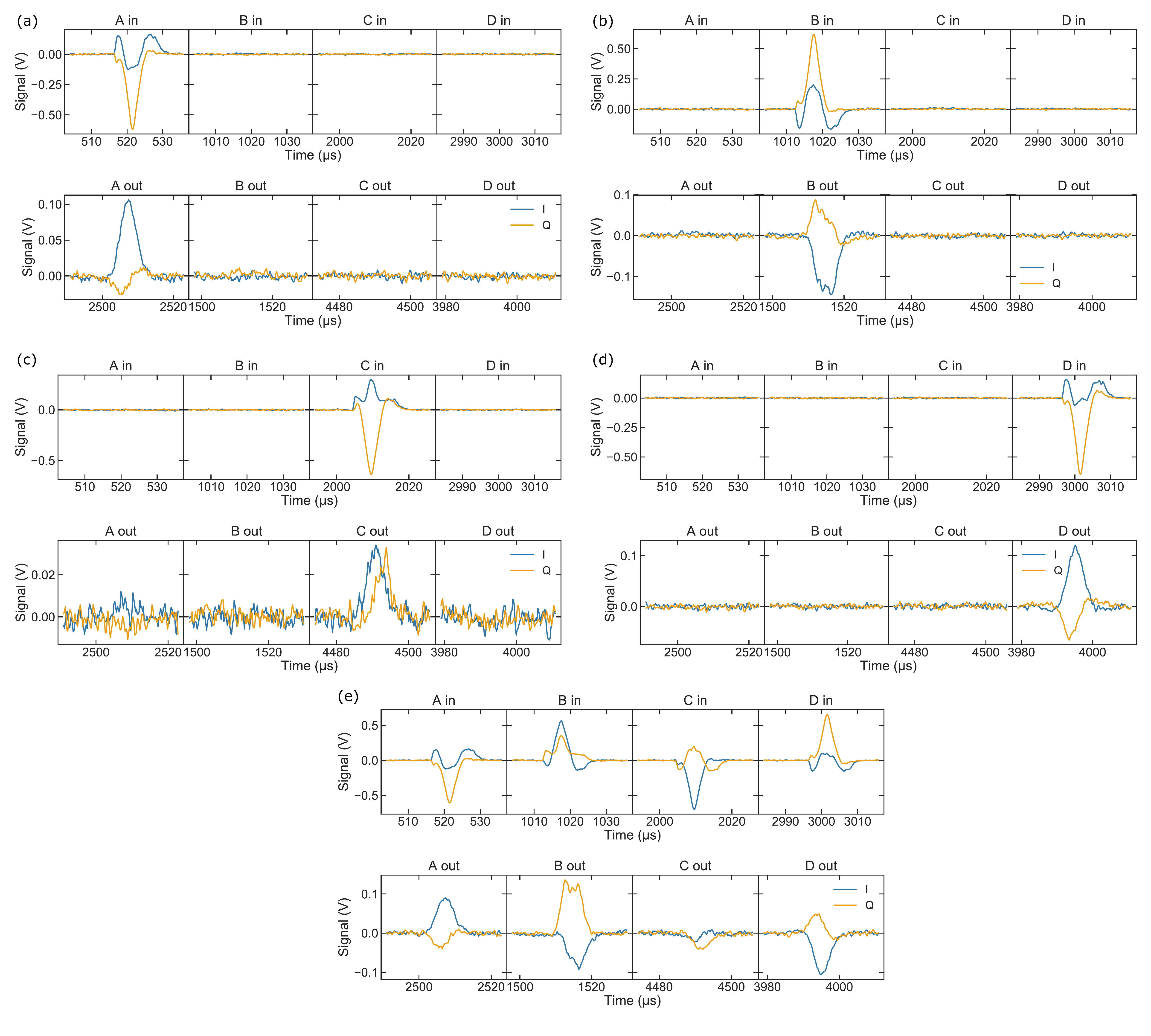}
    \label{fig:highpower}
    \caption{The memory sequence shown in Fig.3 was performed at higher input power ($\sim 15000$~photons per excitation pulse); the raw transient trace of the sequence is shown in (e). In  Figs.(a-d) the same experiment was repeated with only one of each of the 4 input excitation pulses, A,B,C and D turned on. This allows us to unambiguously determine which echo originates from which excitation pulse.  We see that a single echo appears in the expected read slot in each case. We also see that no other echoes appear when only one excitation is turned on. In this power regime, as well as amplifying (non-linearly), the JPA also imparts poorly characterized phase shifts seen particularly when comparing (c,d,e).}
\end{figure}

To further confirm the memory protocol functions as intended, we ran identical memory protocols cycling different excitations on and off so that in each run only one excitation was used. To speed up this measurement we increased the power of the input pulses and the results are shown in Fig.~\ref{fig:highpower} where we also show a replication of the full memory protocol at higher power. In each of the sequences where there is only one excitation stored, we only retrieve one echo, and the echo occurs when we would expect the echo to form further confirming that the memory protocol works.

\bibliographystyle{unsrt}
\bibliography{bibliography}